\documentclass[12pt.preprint]{aastex}

\def\h2o{H$_2$O}
\def\co{$^{12}$CO}
\def\13co{$^{13}$CO}
\def\kms{km s$^{-1}$}

\begin{document}

\title{SWAS Observations of Water in Molecular Outflows}
\author{Jonathan Franklin\altaffilmark{1}, Ronald L. Snell\altaffilmark{1}, 
Michael J. Kaufman\altaffilmark{2}, Gary J. Melnick\altaffilmark{3}, 
David A. Neufeld\altaffilmark{4}, David J.Hollenbach\altaffilmark{5} 
and Edwin A. Bergin\altaffilmark{6}}

\altaffiltext{1} {Department of Astronomy, LGRT 619, University of Massachusetts,
710 North Pleasant Street, Amherst, MA 01003}
\altaffiltext{2} {Department of Physics, San Jose State University, One Washington
Square, San Jose, CA 95192}
\altaffiltext{3} {Harvard-Smithsonian Center for Astrophysics, 60 Garden Street,
Cambridge MA 02138}
\altaffiltext{4} {Department of Physics and Astronomy, Johns Hopkins University,
3400 North Charles Street, Baltimore, MD 21218}
\altaffiltext{5} {NASA Ames Research Center, Moffett Field, CA 94035}
\altaffiltext{6} {Department of Astronomy, University of Michigan, 825 Dennison
Building, Ann Arbor, MI 48109}

\begin{abstract}

We present detections of the ground-state $1_{10}\rightarrow1_{01}$ transition 
of ortho-\h2o\ at 557 GHz in 18 molecular outflows based on data from the {\it Submillimeter 
Wave Astronomy Satellite} (SWAS).  These results are combined with ground-based observations
of the J=1-0 transitions of $^{12}$CO and $^{13}$CO obtained at the {\it Five College
Radio Astronomy Observatory} (FCRAO).  Data 
from {\it Infrared Space Observatory} (ISO) for a subset of the outflows
are also discussed.  Assuming the SWAS water line emission originates from the same
gas traced by CO emission, we find that the outflowing gas in most outflows  
has an ortho-\h2o\ abundance relative to H$_2$ of between about 10$^{-7}$ and 10$^{-6}$.  
Analysis of the water abundance 
as a function of outflow velocity reveals a strong dependence.  The abundance of
ortho-\h2o\ increases with velocity and at the highest outflow velocities some of the
outflows have relative ortho-\h2o\ 
abundances of order 10$^{-4}$.  However the mass of very high velocity gas with 
such elevated \h2o\ abundances represents less that 1\% of the total outflow gas mass.  
The ISO LWS observations of high-J rotational lines of CO and 
the 179.5 $\mu$m transition of ortho-\h2o\ provide evidence for a warmer
outflow component than required to produce either the SWAS or FCRAO lines.  
The ISO line flux ratios can be reproduced with C-shock models with shock velocities 
of order 25 km s$^{-1}$ and preshock densities of order 10$^5$ cm$^{-3}$; these C-shocks 
have post-shock relative water abundances greater than 10$^{-4}$.
The mass associated with the ISO emission is also quite small compared with the total
outflow mass, and is 
similar to that responsible for the highest velocity  water emission detected by
SWAS.  Although the gas responsible for the ISO emission has elevated levels 
of water, the bulk of the outflowing gas has an abundance of ortho-\h2o\ well
below what would be expected if the gas has passed through a C-shock with shock 
velocities greater than 10 km s$^{-1}$.  Gas-phase water can be depleted in the 
post-shock gas due to freeze-out onto grain mantles, however the rate of freeze-out is 
too slow to explain our results.  Therefore we believe that only a small fraction 
of the outflowing molecular gas has passed through shocks strong enough to fully 
convert the gas-phase oxygen to water.  This result has implications for the 
acceleration mechanism of the molecular gas in these outflows.

\end{abstract}


\section{Introduction}

Stellar winds are believed to play a pivotal role in the process of star formation.
During the accretion phase of star formation, stellar winds carry away excess angular 
momentum that would otherwise prevent further collapse.  The winds interact with and 
accelerate the surrounding medium, and as a result, drive molecular outflows with masses 
often much greater than that of the young star itself.  However, the exact mechanism by which
these winds interact with and accelerate the surrounding medium is still a matter of debate. 
A number of different models have been proposed including jets 
with bow shocks, jets with turbulent entrainment, and wide angle 
winds \citep{arce02}.  The various models result in different spatial and 
velocity distributions of the entrained material, and therefore one approach to
distinguish between the proposed models is to map the spatial and velocity structure 
of the outflowing molecular gas \citep{arce06}.  Alternatively, the various acceleration 
mechanisms heat the gas in different ways, resulting in changes in the chemistry of the 
outflow as ices are sublimated and endothermic reactions pathways are opened.  Thus,
determinations of the chemical abundances of the outflowing molecular gas
may permit us to distinguish between the various mechanisms.  The review by \citet{arce06}
suggested that molecules such as SiO, CH$_3$OH, H$_2$O and sulfur-bearing molecules 
have their abundances affected by outflow activity.  For example, \citet{bach01} study of 
L1157 revealed significant abundance variations and suggested that the chemistry of the
outflows may be useful in understanding the time evolution of outflows.

The abundance of water can be strongly affected by shocks \citep{kauf96,berg98} 
such as those predicted to exist in outflow regions.  Moderate velocity shocks  
are capable of producing large abundances of water by releasing 
frozen water from dust grains and by driving all free oxygen into water through 
a series of gas phase chemical reactions \citep{kauf96}.  Thus, water should be a good
tracer of shocks.  The chemical reactions that produce water
have been thoroughly modeled and it has been shown that the production mechanism 
and resulting abundance is very sensitive to temperature \citep{elit78a,elit78b}.  
At gas temperatures below $\sim$300 K water is formed most readily through a series of ion 
molecule reactions that lead to the formation of H$_3$O$^+$.  
The dissociative recombination of H$_3$O$^+$ has several possible outcomes, but has 
been measured in the lab to produce water in  
approximately 25\% of the interactions \citep{jens00}.  This process of water production 
is relatively slow  and modelling has shown
that in a quiescent medium of density 10$^5$ cm$^{-3}$ and temperature 30 K, a water 
abundance of roughly 10$^{-7}$ relative to molecular hydrogen is achieved after 
10$^5$ years \citep{berg98}.

Once the temperature of the gas rises above $\sim$300 K a series of endothermic neutral-neutral
reactions are activated.  At elevated temperatures, these reactions are 
very rapid compared to the ion molecule reactions, and all free oxygen is quickly 
driven into water \citep{elit78a,elit78b}.  \citet{kauf96} show that the passage of a moderate 
C-shock with a velocity greater than $\sim$10 {\kms} is able to elevate temperatures above 300 K 
long enough to produce greatly enhanced water abundances on the order of $\sim10^{-4}$ 
relative to H$_2$.  More importantly, \citet{berg98} showed that 
the enhanced water abundance persists in the post-shock gas after the gas has cooled 
and to significantly reduce the
water abundance required  $\sim10^5$ years even at densities as high as 10$^5$ cm$^{-3}$. 
Since outflow ages are typically estimated to be of order 10$^5$ years or less, any of the
outflowing molecular gas that has passed through a moderate shock should be imprinted
with this greatly elevated water abundance.  After $\sim10^5$ years the abundance of 
water will drop as the water chemistry comes into equilibrium at the lower 
gas temperature and water freezes onto the cold dust grains.  Evidence for this enhancement
is quite apparent in the water observations of some outflows 
such as in Orion-BN/KL \citep{harw98,wrig00,meln00b}.

The ideal nature of water as a tracer of shocked regions is countered by 
the difficulty of detection from ground based observatories.  This 
problem was overcome by the successful launch of three satellites, the {\it Infrared Space
Observatory} (ISO), the {\it Submillimeter Wave Astronomy Satellite} (SWAS) and {\it Odin}.  
ISO was capable of detecting a number of water transitions excited in gas warmer than 
about 80 K \citep{cleg96} while SWAS \citep{meln00a} and {\it Odin} \citep{nord03, hjal03}
can observe the fundamental ortho-{\h2o} $1_{10}-1_{01}$ transition at 538.3 $\mu$m.  
More recently, the {\it Spitzer Space Telescope} has detected several mid-infrared 
water transitions toward NGC 2071 \citep{meln07}. The fundamental transition of ortho-\h2o\
can be readily excited at temperatures greater than about 20 K,
permitting SWAS and {\it Odin} to trace out much cooler post-shock gas than ISO
or {\it Spitzer}.  SWAS has detected ortho-{\h2o} emission associated with the quiescent dense 
molecular gas in numerous cloud cores.  The relative abundance of 
ortho-{\h2o} to H$_2$ in these regions was found to be on the order of 
$\sim$10$^{-9}$ to 10$^{-8}$ \citep{snel00b}, roughly 2 orders of magnitude 
less abundant than predicted by quiescient gas phase chemistry.  
The discrepency is likely the result of water ice mantle formation on the 
dust grains \citep{berg00}. 

In this paper, we present SWAS detections of {\h2o} emission from 18 well studied
outflow regions.  We selected nearby outflows (within 1 Kpc) with significant SWAS
observations that we were able to obtain additional mapping
data from the {\it Five College Radio Astronomy Observatory} (FCRAO).
These outflows are driven by a mix of high and low luminosity Young Stellar Objects (YSOs).  
Eight of these outflows (L1448-mm, NGC 1333-SVS13, HH25mm, $\rho$ Oph A (VLA1623), 
L1689N, Ser SMM1, L1157, L1228) are driven by relatively low luminosity YSOs that
are in most cases individual low mass stars.  The other ten sources (GL490, Orion KL, OMC 2, 
NGC2071, MonR2, NGC2264 D, NGC2264 C, IC 1396N, S140, Ceph A HW2) are driven by 
much higher luminosity YSOs that are associated with newly forming groups and clusters.  
We combine the SWAS {\h2o} detections with maps of {\co} and {\13co} emission 
obtained at the  FCRAO 14-m telescope to determine the water abundance 
in the outflowing gas.  Additionally, we use published ISO results for six of these
outflows to further expand our understanding of the SWAS results.

\section{Observations}

\subsection{SWAS}
      
SWAS was a NASA Small Explorer Mission 
that operated successfully from 1998 to 2005 \citep{toll04}.  SWAS simultaneously
observed the $1_{10}-1_{01}$ transition of ortho-{\h2o} at 556.936 GHz, the
J = 5-4 transition of {\13co} at 550.926 GHz, the 3,1-3,2 transition of O$_2$ at
487.249 GHz, and the $^3P_1-^3P_0$ transition of [CI] at 492.161 GHz.  
O$_2$ emission was not detected toward these outflows and these observations
will not be discussed further.  The high spectral resolution of 
SWAS ($\sim$0.6 {\kms}) permits the kinematic separation of the outflowing gas 
from the ambient material within star forming regions.  However, a detailed 
analysis of the distribution of emission in these outflows is
hindered by the limited angular resolution afforded by the small aperture of
SWAS. The SWAS beam is elliptical, and at the frequency of 
the {\h2o} and {\13co} transitions has angular dimensions of 3$'$.3 $\times$ 4$'$.5
and at the frequency of the O$_2$ and [CI] transitions has angular dimensions of
3$'$.5 $\times$ 5$'$.0 \citep{meln00a}.  Only one pointing 
was obtained for each source; however, in most outflows the large beam encloses the 
entire region of outflow activity.

Observations of these outflows were obtained 
by SWAS over the entire period of mission operations.  The positions observed for the 
18 outflows are listed in Table 1. The data was acquired by nodding the satellite 
alternately between the source position and a reference position relatively free of 
$^{12}$CO emission.  Each position was observed over many orbits and the total integration 
time on source is listed in Table 1.  For each observation, an equal time was spent 
integrating off source toward the reference position.  The on-source integration time
varied from about 6 to 90 hours and thus, the rms noise level achieved varied from source
to source.  The shortest integration time was for OMC2, where the rms noise measured
in the baseline for the H$_2$O and $^{13}$CO spectra was 0.028 K.  The noise
decreased as the square root of time and thus for one of the longer
integration sources, $\rho$ Oph A, an rms of 0.008 K was obtained.  For the source with
the longest integration time, GL490, the baselines were poorer 
and the rms noise obtained was only 0.010 K.  Details concerning data 
acquisition, calibration, and reduction with SWAS are presented in \citet{meln00a}
and \citet{toll04}. 
The data shown in this paper are not corrected for the measured SWAS main beam 
efficiency of 0.90 \citep{meln00a}; however, this correction has been applied 
in our analysis.

\subsection{FCRAO} 

During the fall of 2003 and the spring of 2004, the FCRAO 14-m telescope was used
to obtain maps of the {\co} emission at a rest frequency of 115.271 GHz and 
the {\13co} emission at a rest frequency of 110.201 GHz.  The map centers were the 
same positions as those used for the SWAS observations.  For all observations, the 
32-pixel SEQUOIA array receiver \citep{eric99} was used and data was 
obtained using an On-The-Fly observing technique.  For each source the observations 
were resampled to form maps approximately 6$'$.0 $\times$ 6$'$.0 in $\Delta\alpha$ 
and $\Delta\delta$ in extent with data spaced by 25$''$.  The spectrometer for each 
pixel was a digital autocorrelator with a bandwidth of 50 MHz and 1024 spectral 
channels per pixel leading to a channel spacing of approximately 0.13 {\kms} for 
both {\co} and {\13co}.  The full-width-at-half-maximum beam size of the FCRAO 
telescope at the {\co} frequency is 45$''$.  The spectra in these maps typically 
have an rms noise measured in the baseline of 0.06 K 
for $^{13}$CO and 0.17 K for $^{12}$CO.  The
two exceptions are L1689N and $\rho$ Oph A which transit at low elevation at
the latitude of the FCRAO site and had an rms noise measured in the baseline
of only 0.14 K for $^{13}$CO and 0.47 K for $^{12}$CO.
The CO data shown in this paper
have not been corrected for main beam efficiency, estimated to be 0.45 at the 
frequency of the {\co} line and 0.49 at the frequency of the $^{13}$CO line.  
These corrections have been applied in all subsequent analysis.

\subsection{ISO}

ISO had a much greater wavelength coverage than SWAS and was capable of measuring
the far-infrared emission from a variety of ortho- and para-{\h2o} transitions as 
well as a series of high-J CO emission lines  \citep{cleg96}.  
Published ISO results are available for seven of the outflows in our sample: these
outflows are L1448-mm \citep{nisi99,nisi00,froe02}, NGC1333 SVS13 \citep{moli00}, 
Ceph A HW2 \citep{froe02}, L1157 \citep{neuf00,gian01}, L1689N \citep{cecc98}, 
HH25mm \citep{bene00}, Orion KL \citep{lera06}, and Ser SMM1
\citep{lars02}.  Water lines were detected by ISO in all of these outflows except
Ceph A HW2.  The published line fluxes that we use in this paper were all 
obtained with the LWS instrument.  With the exception of Orion KL, these spectra 
were obtained at relatively 
low spectral resolution (R $\sim$ 300) so they contain
little velocity information.  The lowest lying water line observed by ISO is 
the $2_{12}-1_{01}$ transition at 179.5 $\mu$m. This transition originates from
a higher energy level (E/k = 80 K) than the transition observed by SWAS and
therefore may probe a higher temperature component of water.  ISO 
has a smaller beam than SWAS ($\sim$75'') but the resolution 
is still too poor to permit a detailed spatial analysis.  

\section{Summary of Observations}

In Figure 1, we present the SWAS spectra of the $1_{10}-1_{01}$ transition of 
ortho-{\h2o} for the 18 outflows in our sample.  Each spectrum has only had a 
linear baseline removed.  Seven of the outflows presented in this paper have 
had SWAS data published previously: NGC 2071 and L1157 \citep{neuf00}, 
L1448-mm and IC1396N \citep{bene02}, NGC1333 SVS13 \citep{berg03}, S140 and Mon R2
\citep{boon03}, and L1689N \citep{star04}.  L1157, IC1396N and S140  
have been reobserved by SWAS since the time of these publications and 
the complete spectra are shown in Figure 1.  The initial observations
of GL490 had a ripple in the baseline and so this source was reobserved.  For the new observations
the local oscillator setting was shifted to a value that limited the 
extent of useable spectrum on the blue side of the line.  With the limited 
spectral coverage, placement of the baseline in GL490
is more questionable than in the other outflows, and therefore the extent of the outflow
emission in the blue wing is very uncertain.  Finally, for L1448-mm the position observed 
by SWAS and the center of the CO mapping at FCRAO was offset 
$\sim$3 \arcmin\ north-west of L1448-mm, the center of the large bipolar outflow. 
Thus, our observations of L1448-mm primarily probe the blueshifted lobe of the outflow.
The SWAS results for {\13co} and [CI] will be presented later in the paper.

We have obtained nearly complete CO maps of all 18 outflows.  In only L1228, L1448-mm 
and NGC2264 D does the molecular outflow extend appreciably beyond 
the 6$\arcmin\times6\arcmin$ region mapped. 
The CO maps of HH25mm and Ser SMM1 show obvious evidence of multiple outflows in the
region mapped, and this has been noted previously \citep{gibb93, davi99}.  The SWAS
spectra, although centered on HH25mm and Ser SMM1, include emission from other 
outflows in these regions.  Since all 18 of these outflow regions have been 
extensively studied and there are many published papers with outflow maps, we do not
believe our CO maps provide sufficient new information
to justify showing here.  A summary of the properties of these outflows and their
driving sources are provided in the catalog introduced in \citet{wu04}.

We convolved the CO data to the SWAS resolution to compare \h2o\ and CO line profiles.  
However, since the outflows do not fill the SWAS beam, convolving 
the CO data to match the large SWAS beam substantially
degraded the signal to noise in the high velocity CO line wings making profile comparison
difficult.  So instead, we have chosen to co-average only those {\co} spectra which
have an integrated intensity in either the red or blue outflow wings that is at 
least one-half of the integrated intensity of the peak wing emission.  In these same
positions we have also co-averaged the \13co spectra.  Both \co\ and \13co\ spectra 
averaged in this manner are shown accompanying the water spectra in Figure 1.
It is worth noting a few features in the averaged CO spectra.  First, in Ceph A HW2 there 
is a additional velocity feature at $V_{lsr}$ = 5 \kms\ that is unlikely related 
to the outflow.  In GL490 and L1228 a weak negative feature appears in these spectra,
suggesting that the reference positions were not free of {\co} emission at all velocities.
We excluded the velocity intervals associated with these features in 
the three sources in our subsequent analysis.

A cursory inspection of the CO and \h2o\ line profiles reveal many similarities that
suggest that the emission in these lines likely originates in the same outflow component.
With the exceptions of Mon R2, NGC 2264 D, and S140, the \h2o\ emission 
can be traced to a much higher outflow velocities than
the CO emission.  This effect is most pronounced in the low
luminosity outflows, such as L1157 or HH25mm, and may be due to differences in the
optical depth of these lines.  The CO J=1-0 transition is intrinsically weak, and we
might expect the optical depth of this line to be less than the water line.
Higher rotational transitions of CO, which have larger optical depths, do reveal 
higher velocity emission.  A good example is L1448-mm where obervations 
of the J=2-1 transition of CO \citep{bach90} reveal a velocity extent more similar
to what is seen in the water line.
Although the detection of outflows is more difficult in the J=1-0 transition, the 
smaller optical depths make it ideal for deriving the outflow column density.
We also find that the intensity ratio of the ortho-{\h2o} emission 
to {\co} emission increases with increasing outflow velocity.
Later in the paper we will model these emissions, including the effects of optical depth,
and test whether the observed changes in the intensity ratio is due to variations in
the water abundance with velocity.

Finally we note that nearly all of the {\h2o} spectra show evidence for self-absorption 
similar to that seen in \co.  The self-absorption feature is often much less 
evident in the co-added \co\ spectra shown in Figure 1 than it is in the
individual \co\ spectra obtained toward the center of the outflow.  The self-absorption 
feature seen in both \h2o\ and \co\ presumably arises due to foreground, quiescent, 
low-excitation gas absorbing the central part of the broad emission produced by the outflowing gas.

\section{Analysis and Results}

To derive the water abundance from the SWAS ortho-{\h2o} observations, we must 
quantify the density, temperature and column density of the outflowing 
molecular gas.  Densities and temperatures are extremely difficult to 
determine in molecular outflows and 
few studies have attempted to measured these properties.
Some results are available concerning the gas temperature
in outflows and are summarized in  \citet{wu04}.  However
for the density, it is difficult to find any reliable
estimates.  In this paper we will simply assume values of temperature
and density typical of star forming cores.
Specifically, we set the density to 10$^5$ cm$^{-3}$ and the temperature 
we assume varies from 30-100 K, depending on the luminosity of the
driving source.  For outflows associated with low-luminosity YSOs (L1448-mm, NGC 1333 SVS13,
HH25mm, $\rho$ Oph A, L1689N, Ser SMM1, L1157, L1228 and IC1396N) we assume
a temperature of 30 K.  For outflows associated with more luminous YSOs ( GL490, 
OMC2, NGC2071, MonR2, NGC 2264 D, NGC2264 C, S140 and Ceph A HW2 ) we assume
a temperature of 50 K and for Orion KL we assume a temperature of 100 K.  These
assumptions are an oversimplification of the complex temperature and density
structure likely to be present in these
outflows.  A number of outflows were systematically
investigated by \citet{levr88} and the range of densities and temperatures 
estimated for outflows in this study are consistent with our assumptions.

Higher temperature gas is present in nearly all of the outflows studied here 
as evidenced by the presence of shock excited optical emission (Herbig-Haro objects) as
well as highly excited lines of a variety of molecular species seen at infrared and
far infrared wavelengths.  In particular, ISO has detected high-rotational
transitions of CO and \h2o.  However, the bulk of the outflowing molecular gas likely has 
much lower gas temperatures.  In our analysis we will first address the emission from 
the cooler gas (T = 30-100 K) that we believe predominately gives rise to the emission 
in the lowest rotational transitions of CO.  In \S~4.6 we will address specifically 
the emission from the warmer gas within the outflows.

\subsection{Outflow Gas Column Density}

We derive the column density of outflowing gas based on the {\co} J=1-0 emission  
assuming the outflow gas temperature and density as described in the previous section. 
It is necessary to begin by identifying the velocity
intervals over which emission from the outflowing gas dominates the ambient cloud emission.
As our guide for setting these intervals, we have used the observed \13co\ spectra
and defined the break in line profile shape  as the division between ambient
and outflow emission.  The intervals so defined are given in Table 2 and illustrated
in Figure 1.  Based on these velocity intervals, we compute the integrated
intensity of the CO emission in each map position for each outflow.  The 
emission from the outflowing gas within the velocity interval around the quiescent
cloud velocity is excluded in our analysis; however, this velocity interval is 
relatively small and its exclusion will not strongly affect the results.

The integrated intensity of CO can be converted to a column density, assuming
that the emission is optically thin and in LTE.  At the high densities assumed for the 
molecular outflow gas (much greater than the critical density for {\co}) the 
assumption of LTE is extremely good.  The total CO column density can be written as

\[N=1.15 \times 10^{14}e^{5.54/T}(\frac{1}{3}+0.36T)\int T_{mb}\textrm{d}v,\]

\noindent
where T is the excitation temperature (which is equivalent to the gas temperature in LTE), 
and $T_{mb}$ is the main beam antenna temperature (antenna temperature corrected by 
the main beam efficiency) of the J=1-0 transition of CO.

Although our assumption of LTE is likely valid, we have no reason to believe 
that the observed high velocity {\co} emission is optically thin.  Our simultaneous 
observations of the {\13co} emission from the outflows enable us to test this thin 
assumption and, if necessary, correct for any optical depth effect.  We estimate the
optical depth of this gas by comparing the observed ratio of {\co}/{\13co} emission to the assumed 
local interstellar medium ratio of 65 \citep{lang90}.  Assuming equal excitation temperatures 
for \co\ and \13co, the
relation between the observed isotopic line ratio, R, and the optical depth can be written as 

\[R=\frac{1-e^{-\tau}}{1-e^{-\tau/65}},\]

\noindent
where $\tau$ is the optical depth of the {\co} emission.  If we further assume that 
the {\13co} emission in the high velocity gas is optically thin ($\tau$/65 $\ll$ 1) 
then the ratio becomes \[R \approx \frac{65}{\tau}(1-e^{-\tau}).\]  We then apply a
correction to the column densities derived assuming the emission is optically thin,
and the corrected column density is then simply

\[N_{thick}=N_{thin}\frac{\tau}{1-e^{-\tau}}\approx N_{thin}\frac{65}{R}.\]

The sensitivity of our \13co\ maps is insufficient to detect high velocity gas in 
individual spectra. Because of the limited angular extent of the outflows, if
we average over the entire map or even over the SWAS
beam, we severely reduce the outflow signal in the averaged spectrum. 
Therefore, we have averaged only those spectra 
that show evidence of the outflow in {\co} emission, following the method used to 
construct the {\co} and {\13co} spectra shown in Figure 1.  The redshifted and 
blueshifted gas were considered separately, unlike the method used to produce the
spectra shown in Figure 1, and only those positions with {\co} 
integrated areas greater than one-half the peak value detected 
for the redshifted or blueshifted outflow in that source were selected and averaged 
together.  The {\13co} profiles were created by averaging the same positions 
selected by the {\co} emission.  The observed isotope ratio in either the redshifted
or blueshifted outflow is then measured from these averaged spectra.  The measured 
isotopic ratios are presented in Table 2.  The measured value of R is typically between
20 and 30 which correspond to optical depths of 3.0 to 1.8 and to column density 
corrections of a factor of 3.3 to 2.2.  The average optical depth correction is
used to correct the optically thin derived column density throughout each outflow lobe.

\subsection{{\13co} J=5-4 Emission - Test of Outflow Model}

The SWAS observations of the J=5-4 transition of {\13co} can be used to test our 
adopted phyical properties and derived gas column density of these outflows.  
The emission in this line originates from an upper state with an energy of E/k = 79 K;
thus it is particularly sensitive to the assumed temperature in the range considered
for these outflows.  Significant {\13co} J=5-4 emission was detected in seven of the 
outflows (Ceph A HW2, GL490, MonR2, NGC2071, Orion KL, OMC 2 and S140) all driven by
luminous YSOs.  For outflows associated with lower luminosity YSOs, {\13co} J=5-4
emission was either extremely weak or not detected.  This general trend is consistent
with our assumption that the outflowing gas associated with more luminous YSOs is
warmer than that in the lower luminosity YSOs.  However, a more quantitative test 
of our physical model can be made.  

Using the computed CO gas column density and the assumed temperature and density 
for the outflows, we can predict the emission in the {\13co} J=5-4 transition 
at each location in the outflow where we have determined the outflow column density.  
As in our previous column density determination,
we assume that the abundance ratio of CO to {\13co} is 65.  
We convolve the map of predicted emission with the SWAS beam,
and compute the predicted {\13co} J=5-4 line flux in each of the outflows and compare that
with the line flux observed by SWAS.  We find that for most of the sources with significant
{\13co} J=5-4 emission that the modeled and observed 
line fluxes are consistent within a factor of 2.  The agreement is remarkably good, since 
only small changes in the density and temperature can greatly affect the line flux.  
The most discrepant outflows were Mon R2 and GL490, 
where the modeled line flux was 3 to 6 times larger than that observed.  In
three of the outflows associated with luminous YSOs, IC1396N, NGC2264 C and 
NGC2264 D, there was no detectable {\13co} J=5-4 emission; however, in all cases the
non-detections are consistent with the model predictions.  For the
outflows associated with low luminosity YSOs, only $\rho$ Oph A and L1689N have possible
weak outflow detections in the {\13co} J=5-4 line, and the other sources (L1448-mm, NGC1333-SVS13
SVS3, HH25mm, Ser SMM1, L1157, and L1228) were not detected.  We find again that for
these low-luminosity outflows with weak detections or only upper limits the 
results are consistent in most cases with our model.  The one exception is  
HH25mm, where the modeled flux exceeds the 3$\sigma$ upper limit by a factor 
of approximately 3 to 5.
In all cases where there is poor agreement, the model predictions exceed the observed
line flux suggesting that we have either overestimated the
temperature or density in our outflow model.   

We have used two of the most discrepant outflows, MonR2 and HH25mm, to estimate the magnitude 
to which we may have overestimated either the density or temperature in our
outflow models.  Unfortunately with only the {\13co} J=5-4 data it is 
impossible to derive a unique temperature and density for the outflows, since
both parameters affect the observed line flux.  For these calculations we have recomputed
the gas column density using the modified temperature or density and then recomputed
the predicted {\13co} J=5-4 emission.  In MonR2, we have found
that by reducing the temperature from 50 to 30 K we can obtain good line flux agreement.
However, we can also maintain the temperature at 50 K and reduce the density by a factor
of 5 and obtain good agreement.  Of course the correct
solution could be a combination of both a lower temperature and a lower density.  
Likewise in HH25mm, we can achieve good flux agreement by either lowering the 
temperature from 30 to 20 K or by decreasing the density by a factor of 4.  Nevertheless,
in most outflows the line flux agreement suggests that the temperature is
probably accurate to 20\% and the density accurate to within a factor of two.

\subsection{Mass of Outflows}

The mass of outflowing redshifted and blueshifted gas in each outflow can be
found using our map of the CO column density and
the distances given in Table 1.  We have assumed a CO/H$_2$ ratio of
1$\times$10$^{-4}$ for the mass determinations which are summarized in Table 2.   
The statistical errors on the observed integrated intensity of 
{\co} are small, and the uncertanity in our mass estimates are dominated by the 
uncertainties in our assumed temperature, correction for optical depth, and CO to H$_2$
conversion.  Uncertainties in density have little impact on the mass determination, unless
it is substantially lower that the value we assumed.

Our method for estimating the water abundance requires a knowledge of the
spatial distribution of the column density of outflowing gas.  For this
analysis, which we describe in the next section, we have only included positions 
which have a $3\sigma$ detection of {\co}.  
The mass estimates in Table 2 are based only on positions within the
outflow with CO detections, and thus may underestimate the total outflow mass.  
Statistically we can provide a much better estimate of the total outflow mass by
using all the positions, regardless of whether the integrated intensity is
positive or negative.  We have recomputed the mass of each
outflow without applying a detection threshold, and find masses in the redshifted and
blueshifted outflows that are at most only ten percent larger than those presented in
Table 2.

\subsection{{\h2o} Abundance}

We first estimate the ortho-{\h2o} abundance in the outflows assuming that 
the \h2o\ emission seen by SWAS originates in the same gas as traced by the CO J=1-0 
emission.  Later we will explore whether this is reasonable assumption and if there are
other possibilities for its origin.  The integrated intensity of the 
SWAS \h2o\ line in each outflow has been determined using the same velocity intervals
that were used for CO.  The observed integrated intensity and line flux for the
redshifted and blueshifted emission is given in Table 2.
We model the water emission using 
a statistical equilibrium code that uses the Large Velocity Gradient (LVG) approximation 
to account for radiation trapping.  The LVG assumption should be a valid 
approximation for the observed broad-line emission from these outflows. Collisional 
rate coefficients are taken from \cite{phil96}.  Since the ortho-H$_2$ and para-H$_2$ 
collisional rate coefficients with \h2o\ are different by nearly an order of
magnitude, the assumed ratio of ortho-H$_2$ to para-H$_2$ is important.  We 
have assumed that the ratio of ortho- to para-H$_2$ is in LTE as found for the
outflows studied by \citet{neuf06} .  Thus for our assumed gas temperatures of
30, 50 and 100 K, the ratio of ortho- to para-H$_2$ is approximately 0.03, 0.3
and 1.6, respectively.
We include the five lowest levels of ortho-{\h2o} in our calculation.  

The physical inputs to our \h2o\ model are the temperature, density, and column density 
distribution of the outflowing gas.  We have used the same temperature and density as 
described earlier for computing the CO column density.  
The process of solving for the abundance begins by assuming a water abundance and then 
computing the predicted water emission at each position in the outflow.
We convolve the predicted emission with the SWAS beam, and then compare with 
the observed integrated intensity.  The abundance of ortho-{\h2o} is then varied 
until we achieve agreement between model and observations.  
This approach is very similar to that used by \citet{snel00a} and has the 
advantage of correctly accounting for optical depth variations and beam dilution
effects across the large SWAS beam based on our measured gas column density distribution.  
Since the input to our model is the 
measured CO column density, our analysis provides a direct determination of the 
ortho-{\h2o} abundance relative to CO.  We infer an abundance of 
ortho-{\h2o} relative to H$_2$ by assuming a \co\ to H$_2$ ratio of 
10$^{-4}$. The resulting ortho-{\h2o} abundances relative to H$_2$ are 
presented in Table 2.  

We have investigated the impact on the relative ortho-\h2o\ abundance for different 
assumed temperatures and densities. 
Unless the assumed density is dramatically decreased, density uncertainties have
little impact on the determination of the CO column density.  However, density
does have a big effect on ortho-\h2o\ abundance.  Since the water emission in these
outflows is approximately in the effectively thin limit \citep{neuf00}, the abundance derived
will be inversely proportional to the assumed density.  Thus, if we were to 
assume a density a factor of two larger, the resulting ortho-\h2o\ abundance will 
be a factor of two smaller.  Changes in the assumed temperature,
on the other hand, will effect both the derived column density of CO and 
ortho-{\h2o} abundance.  We
have used the redshifted outflows of MonR2 and L1689N as test cases.  In each case we
have increased and decreased the temperature by a factor of two and recomputed the
CO column density and ortho-\h2o\ abundance.  We found for these outflows that 
increasing the temperature by a factor of two results in a decrease in the relative
abundance of ortho-{\h2o} by a factor of 3 in Mon R2 and by a factor of 8 in L1689N.
Decreasing the temperature by a factor of two results in an increase in the derived
ortho-{\h2o} abundance by a factor of 9 in Mon R2 and by a factor of 5 in L1689N.
Thus, our derived ortho-{\h2o} abundances are very sensitive to the assumed physical 
conditions in the outflowing gas, which is unfortunate since these conditions are 
very poorly known.  Thus, the systematic errors introduced by our poor knowledge
of the outflow density and temperature dominate over the uncertainties in the 
\h2o\ line fluxes in determinations of the water abundance.

The abundance of \h2o\ relative to H$_2$ spans a wide range of values,
from 10$^{-5}$ to 10$^{-8}$.  This range of abundances is so large, we believe
it is unlikely due to only uncertainites in the physical conditions and must reflect
true abundance variations.  We note that the results for NGC2071 and L1157 are in 
good agreement with the abundances reported earlier by  \citet{neuf00} that are also
based on the SWAS data.   We find the highest \h2o\ abundances are found in 
Orion KL and L1157, and the lowest abundance in Mon R2.  Our derived ortho-{\h2o} 
abundances in nearly all of the outflows are much larger than that derived for
quiescent molecular cloud gas \cite[$\sim$10$^{-9}$;][]{snel00b} which suggests 
some enhancement of the water abundance in outflows.  
The largest abundance derived, $\sim$10$^{-5}$ in Orion KL, is still less
than one might expect if all of the free oxygen had been coverted into water
as is expected for a moderate shock.  It is very important 
to remember that these abundance estimates are averages over 
the entire SWAS beam, and do not rule out the possibility of much higher abundances
existing over a small portion of these outflows.
In fact, in the warm gas probed by ISO, the abundance of \h2o\ is 
as high as 5$\times$10$^{-4}$
in the warm gas in Orion KL and L1448-mm \citep{harw98, gian01}.  The relationship
of the warm gas probed by ISO and the emission seen by SWAS will be examined in 
more detail in \S~4.6.

\subsection{Velocity Dependance of {\h2o} Abundance}

The high spectral resolution of SWAS enables us to further investigate the 
abundance of {\h2o} as a function of velocity.  
In the previous section we assumed that the ortho-{\h2o} abundance is 
constant thoughout the outflow.  We will now investigate whether the abundance 
of ortho-{\h2o} relative to {\co} varies with outflow velocity.

We have divided the redshifted and blueshifted emission from the outflowing
gas in CO and \h2o\ into velocity bins, each of width 5 {\kms}, continuing 
to higher velocities until {\co} is not detected.  For a velocity interval 
to be analyzed, at least two positions within the outflow must have significant 
detections of {\co} emission.  The integrated \h2o\  intensity and line flux 
as a function of velocity are presented in Table 3 for each outflow.
Optical depth corrections were considered separately for each velocity interval 
using the same averaged CO spectra as described before.  Any interval 
with a non-detection of {\13co} was assumed to be optically thin, which 
is often the case at the higher velocities in the outflows.  
We then used the same technique as discussed above to model each of the 
individual velocity intervals.  The results of our analysis are presented
in Table 3, which includes the observed CO isotopic ratio, the mass of gas, and
the ortho-\h2o\ abundance as a function of velocity.  

Our abundance analysis requires \co\ detections.  Although, earlier we suggested
that including only 3 $\sigma$ detections had only small effect on the total 
outflow mass, the impact of truncation is more severe 
in the higher velocity bins where the signal to noise of the \co\ emission
is relatively low.  We have recomputed the mass in each velocity bin without
applying a detection threshold, and found that in the most discrepant
bins, which are at the highest velocities, the mass could be underestimated 
by a factor of 2 to 3.  Thus, the ortho-\h2o\ abundances quoted may by overestimated
by a similar factor.  Despite these uncertainties, we find that in
every outflow the derived abundance of ortho-{\h2o} increases steadily with 
increasing outflow velocity.  Even outflows that did not show obvious disparate 
velocity extents between CO and \h2o\ show ortho-\h2o\ abundance variations 
with velocity.  Some of the most pronouced abundance variations with velocity
are in the outflows driven by low-luminosity YSOs; for example in NGC1333-SVS13 
the ortho-\h2o\ abundance in both the
redshifted and blueshifted high velocity gas increases by two orders of magnitude
from the lowest velocity outflowing gas to the highest velocity outflowing gas.
At the highest outflow velocities, the abundance of ortho-\h2o\ relative to H$_2$ 
gas can be as high as $1\times10^{-4}$, approaching values that  
one might expect if all of the oxygen not in CO is converted into \h2o as 
anticipated if the gas has recently passed through a moderate velocity shock. 
However, since most of the outflow mass is at relatively low outflow velocities, 
the mass of gas with this highly elevated ortho-{\h2o} abundance represents less 
than 1\% of the total mass of outflowing gas, although, it accounts for a much
larger fraction of total water line flux.

\subsection{ISO Analysis}

ISO with the LWS detected both high-J CO lines and a number of 
ortho- and para-{\h2o} lines in eight of the molecular outflows in our survey.  
The lines observed by ISO likely arise in a much warmer 
gas than is responsible for the bulk of the low-J CO outflow emission that we 
have modeled.  For water, the lowest excitation line observed by ISO 
(the $2_{12}-1_{01}$ transition at 179.5 $\mu$m) has a flux in most 
sources more than 10 times larger than that predicted
by the outflow models discussed in previous sections.  Thus, the ISO water
emission must be produced by a warmer, likely shock-excited component, of 
the outflowing gas.  Many authors \citep{harw98, cecc98,nisi00, moli00,
bene00, gian01} have suggested that most of the ISO emission comes
from non-dissociative C-shocks with shock velocities of order 15 to 25 km s$^{-1}$
and pre-shock densities of order 10$^4$ to 10$^5$ cm$^{-3}$. A significant 
enhancement of water in these post-shocked regions is needed to explain the ISO
observations.  Many note that the strong [OI] emission detected in all of these outflows
and the detection of [SiII] in some of the outflows, may require the 
presence of dissociating J-shocks.
\citet{bene02} compared the water emission observed by SWAS 
and ISO in four outflows and suggested that warm, approximately 1000 K, gas
was responsible for the ISO emission and some of the SWAS emission, however they also
concluded that SWAS was also sensitive to a cooler gas component not traced by ISO.

We have used the published line fluxes for L1448-mm \citep{nisi99,nisi00}, 
NGC1333-SVS13 \citep{moli00}, L1157 \citep{gian01}, L1689N \citep{cecc98}, 
HH25mm \citep{bene00}, Ser SMM1 \citep{lars02} and Orion KL \citep{lera06}
to perform a simple C-shock model analysis.  We note that \citet{froe02} 
reported two pointings of ISO toward the Ceph A HW2 outflow; however, \h2o\ emission 
was not detected, so we have not included this source in our analysis.
Our modeling is aimed at estimating the shock solid angle responsible for producing the
ISO emission and whether any significant fraction of the SWAS water line flux
originates in this shocked gas.
In Table 4 we present the flux in the 179.5 $\mu$m line of ortho-\h2o\ observed
by ISO for the six outflows with detected emission that were in common with our survey.  
In L1448-mm, NGC1333-SVS13, and L1157 multiple positions were observed by ISO and we have 
combined the fluxes for positions that are within the SWAS beam.  
In addition to CO and \h2o, ISO detected relatively strong 63 $\mu$m [OI]
emission in these outflows \citep{cecc98,nisi00,bene00,moli00,gian01,lars02}.

To analyze the ISO emission, we use the C-shock models of \citet{kauf96}.  
We have integrated the line intensity contributions in 
the post-shock gas of these C-shock models to the point 
where the gas has cooled to a temperature of 50 K.  
At that point the gas is too cool to contribute much intensity to the observed ISO 
lines and the freeze out of water onto dust grains may become important, which is
not included in this model.  In addition, the
postshock gas has become indistingishable from the gas previously modeled.  
We can use the high-J CO and \h2o\ lines in conjunction with these
C-shock model results to constrain the shock velocity and pre-shock gas density.

The flux in the high-J CO lines and in the 
179.5 $\mu$m transition of ortho-\h2o\  are used to establish the
correct C-shock model.  The intensity ratio of the high-J CO lines, specifically
the ratio of the J=20-19 transition to J=15-14 transition, is very 
sensitive to the temperature of the post-shock gas, and therefore 
provides a constraint on the shock velocity.  The ratio of the 
CO J=15-14 line flux to 179.5 $\mu$m water line flux is sensitive to 
pre-shock density.  Although these line ratios vary somewhat from outflow
to outflow, all outflows require 
shock velocities $\ga20$ \kms\ to produce the observed ratio of CO lines.  For 
example the average CO line flux ratio of the J=20-19/J=15-14 transitions 
observed is 0.45, and this ratio requires a shock velocity of 40 \kms\ for 
a pre-shock gas density of 10$^4$ cm$^{-3}$ or 20 \kms\ for a pre-shock gas 
density of 10$^5$ cm$^{-3}$.  To reproduce the observed flux ratio of CO J=15-14 
to 179.5 $\mu$m \h2o, typically about unity, requires 
post-shock densities of $\ga$ 10$^5$ cm$^{-3}$.  We have adopted a C-shock model 
with a shock velocity of 25 \kms\ and a pre-shock density of 10$^5$ cm$^{-3}$.  
The column of post-shock H$_2$ gas in this C-shock model to
the point where the gas has cooled to 50 K is $6\times10^{21}$ cm$^{-2}$ and 
the integrated intensity of the 179.5 $\mu$m \h2o\ line produced by this column 
of post-shock gas is $8.1\times10^{-3}$ ergs s$^{-1}$ cm$^{-2}$ sr$^{-1}$.  
The \h2o\ abundance relative to H$_2$ in the post-shock gas is $4\times10^{-4}$.  

Based on this adopted C-shock model we estimate the total solid 
angle of shock front needed to reproduce the observe line flux and 
consequently the total mass of shocked gas; these results are given in Table 4.  
With the exception of Orion KL, the angular area of these shocked regions is only 
$\sim$10 square arcseconds, much smaller that the angular area of the outflow 
mapped in the low-J CO lines or the SWAS and ISO beamsizes.  However for Orion KL,
the angular size is about 1280 square arcseconds, comparable to the angular size
of the outflow mapped in the low-J rotational lines of CO \citep{rodr99}.
Recently, \citet{cern06} modeled over 70 far-infrared rotational lines of
water observed by ISO with the LWS.  They modeled the outflow
as a 40$''$ diameter shell expanding at 25 km s$^{-1}$ with a
temperature between 80-100 K, density of 2.5-3.5$\times10^5$ cm$^{-3}$ and 
a relative abundance of water of 2-3$\times10^{-5}$, smaller than that
derived by \citet{harw98}.  The results of \citet{cern06} are similar to
what we quoted earlier based on the SWAS data.  Unlike the other outflows, the 
water emission observed by ISO in Orion KL may arise from the same gas that
is responsible for producing the low-J CO emission.  

With exception of Orion KL, the mass of warm gas needed to reproduce the
ISO observed lines is only 0.0008 to 0.005 solar masses.  Our mass estimates 
are similar to those reported by \citet{bene02} based on a thermal model with 
temperatures between 600 and 1400 K.
Thus, the mass of warm gas responsible for producing the ISO lines is only about 1\% of 
the total outflowing gas mass inferred from our earlier analysis of the J=1-0 \co\ line.  
For Orion KL, the mass in the shell model presented by \citet{cern06} is of order
6 solar masses, similar to our total mass estimate from the low-J CO lines.

These warm, postshock gas regions probed by ISO also produce emission in the 
538 $\mu$m line observed by SWAS.  The adopted C-shock model has a flux
ratio of the 179 $\mu$m/538 $\mu$m lines of ortho-\h2o\ of approximately 32.  
The observed flux ratio of these lines is summarized in Table 4.
It is important to remember that the ISO observations
are made in a smaller beam than SWAS ($\sim$75''), so that a single
ISO observation may not detect all of the 179 $\mu$m emission that is
contained within a SWAS beam.  Even in the outflows with multiple ISO
pointings, the entire SWAS beam is not fully sampled, thus the observed
line flux ratios in Table 4 are only lower limits.  We estimate that the 
fraction of the SWAS water line flux that originates in the warm, post-shock
gas to be as little as 10\% in some outflows to more than 50 \% in others.
\citet{bene02} reached a similar conclusion.  It may be significant that the 
two outflows with the most complete ISO coverage (L1448-mm and Orion KL) have
the largest line flux ratios.  A significant fraction
of the SWAS water line flux may be arising from the very small mass of
warm post-shock gas.

\subsection{Abundance of [CI] in Outflows}

SWAS obtained very high signal to noise spectra of [CI] toward these outflow sources.  
Obvious high velocity [CI] emission was detected in Mon R2, L1228, Ceph A HW2,
Orion KL, and S140 and spectra of [CI] in these sources are shown in Figure 2.
Although no high velocity emission was detected in L1448-mm, the baseline in this
source was somewhat worse.  Also in Ser SMM1, the blue wing of the [CI] line
was contaminated by emission in the reference position.
The presence of high velocity [CI] emission in many of these outflows
was previously noted by \citet{walk93} and \citet{minc94}.

In Orion KL, the spectrum shown in Figure 2 appears to have an additional emission 
feature at a velocity of about 45 km s$^{-1}$.  It is unlikely that this feature
is related to the redshifted outflowing gas, as the line profile would have to
be significantly different from those observed in either CO or \h2o.  
The line is unlikely to arise from the opposite sideband.  Since SWAS does not 
doppler track in real time, frequency corrections are applied to each short time 
segment in the data pipeline, and thus lines in 
the opposite sideband should not add coherently to produce such a discrete feature.
Assuming this feature is in the
same sideband as [CI] and originates in gas with a V$_{LSR}$ of 9 km s$^{-1}$, then
the line frequency is approximately 492.10 GHz.   We searched the JPL line list 
for relatively low excitation lines of molecules known to be present in
Orion and the only candidates are the 13$_{3,10}$-13$_{2,11}$ transition of 
CH$_3$CHO-e at 492.093 GHz and the 13$_{8,6}$-13$_{7,6}$ transition of 
C$_2$H$_5$OH at 492.086 GHz.  The lines of CH$_3$CHO are extremely weak
in the spectrum of Orion \citep{sut85, ziu93}, and thus is probably an
unlikely identification for the feature seen by SWAS.  Ethanol, on the
other hand 
has a number of lines detected in the 336 to 340 GHz range at several locations 
in Orion \citep{sut95}, but the frequency of the transition is significantly
lower than is needed to explain the SWAS feature.
None of the previously published spectra of [CI] in Orion KL had adequate 
sensitivity to confirm or refute this feature.  

Using the same velocity intervals as defined earlier for the redshifted and
blueshifted emission in each outflow, we have computed the [CI] outflow
line flux.  A summary of the [CI] line flux is presented in Table 5.  We note
that we have included the feature seen within the redshifted emission in 
Orion KL as part of the integrated intensity of the outflow.  Although we
have [CI] detections in these velocity intervals for most of the outflow
sources, in many cases the emission may be dominated by the 
gaussian wings of the quiescent line emission.
We computed the CI abundance in an identical manner as we computed the
\h2o\ abundance, using the same physical model for each of the outflows.
The resulting CI abundance relative to H$_2$ is summarized in Table 5.

We find that the abundance of CI relative to H$_2$ is typically about
1$\times10^{-5}$.  The largest abundance of CI is found for NGC2264 D
and $\rho$ Oph A.  However, in NGC2264 D it is questionable whether the weak [CI] 
emission in the red or blue shifted velocity intervals is related to 
the outflow.  Very weak (T$_A^*$ $\sim$ 0.05 K) and very broad [CI] 
emission is detected toward $\rho$ Oph A, similar to the emission profile 
seen in \h2o.  The CI abundance in $\rho$ Oph A is about five times 
larger than in the other outflow.  
For quiescent cloud emission the abundance of CI relative to CO is 
typically 0.1 to 0.5 \citep{zmui88, plum00, howe00}, similar to that
determined for the outflowing gas if we assume a CO to H$_2$
ratio of 1$\times10^{-4}$.  Thus the CI in the outflowing
gas is similar in abundance to that of the ambient gas in agreement
with the results of \citet{walk93}. 
The [CI] spectra have insufficient signal to
noise to investigate variations of the CI abundance with outflow
velocity.

\section{Discussion and Conclusions}

The 1$_{10} - 1_{01}$ transition of ortho-\h2o\ at 538 $\mu$m has been detected 
by SWAS in 18 molecular outflows.  The \h2o\ line profiles are similar
to the line profiles observed for the J=1-0 transition of \co\ and suggest 
that the emission seen in both species may be produced by the same gas.  
If we assume that the SWAS \h2o\ emission arises in the same gas that makes up the bulk
of the molecular outflow, we find that the outflowing gas has 
an ortho-\h2o\ abundance relative to H$_2$ typically between 
10$^{-6}$ and 10$^{-7}$.  However, there are a few exceptions: most notably 
Orion KL and L1157 have anomalously high relative ortho-\h2o\ abundances of 
about 10$^{-5}$, and Mon R2 has an anomalously low relative abundance
of about 10$^{-8}$.  The relative ortho-\h2o\ abundances in Table 2 have substantial
uncertainty that arise almost solely due to the sensitivity of the derived 
water abundances to the assumed temperature and density of the outflowing gas, both of
which are poorly known.  The {\13co} J=5-4 observations by SWAS suggest that
our physical model cannot be too much in error. and thus we believe that the 
abundance of water in most of these outflows is elevated relative 
to that measured in quiescent cloud gas.   We also derive the CI abundance 
in the outflowing gas, and find values that are similar to quiescent cloud material, 
and thus unlike water, the abundance of atomic carbon
appears to be unaffected by the outflow activity.

We also analyzed the velocity dependence of \h2o, and find 
that the abundance of ortho-\h2o\  varies significantly
with velocity.  In nearly all of the outflows, we find a steady increase in the 
ortho-\h2o\ abundance with increasing radial velocity of the outflowing
gas.  In the most striking examples (NGC 1333 SVS13 and HH25mm), the abundance
of water increases by two orders of magnitude from the lowest velocity
to the highest velocity outflowing gas.  Nevertheless, the mass of outflowing gas
with greatly elevated \h2o\ abundance is very small, representing at most
only  $\sim$1\% of the total outflow gas mass.

The ISO observations provide evidence for the presence of a much warmer
outflow component in those outflows observed.  With the exception of Orion KL,
our C-shock modeling of the ISO emission suggest that this warm gas 
constitutes less than 1 \% of the total outflow mass 
and arises from an extremely small fraction of the molecular outflow solid angle.  
However, this gas is expected to have a relative 
abundance of water in excess of 10$^{-4}$.
The presence of this warmer component adds an additional complexity to abundance
determinations in the cooler gas.  Our shock modeling, although simple, suggests
that a significant fraction of the water line flux observed by SWAS could arise in 
this warmer gas component.  Unfortunately the ISO observations did not
cover the full extent of the SWAS beam nor did it have velocity resolution, so
it is extremely uncertain what fraction of the SWAS line flux is associated with
this warm gas probed by ISO.  We estimate that for the outflows with both
ISO and SWAS observations, as much as 50 \%\ of the
SWAS flux could originate in the warmer gas component.
If any significant fraction of the SWAS \h2o\ line flux arises from the warm gas,
then we have overestimated the line flux from the cooler gas traced by the
low-J transitions of CO, and consequently overestimated the ortho-\h2o\ abundance 
in this cool gas.  Therefore our estimates of the water abundance
presented in Table 2 are likely to be too large.  The extent to which the abundance
is overestimated is very uncertain, but if only half of the SWAS line flux is
due to the warm gas this only reduces the \h2o\ abundance by a factor
of two, which is well below other uncertainties in our abundance determination.
Thus, unless nearly all of the SWAS line flux is due to this small mass of
warm, shocked gas, the ortho-\h2o\ abundance in the low velocity
outflowing gas is enhanced relative to that measured for quiescent cloud gas
(relative abundance of only 10$^{-9}$ to 10$^{-8}$ according to \citet{snel00b}).

The studies by \citet{kauf96} and \citet{berg98} demonstrated that even a 
mild C-shock with shock velocities greater than 10 km s$^{-1}$ will elevate the
gas temperature to permit the efficient conversion of 
oxygen into water and to vaporize a large fraction of the water ice on grains.
Thus, most of the oxygen, not in CO, will be rapidly converted 
into water resulting in relative water abundances in excess of 10$^{-4}$.  
\citet{berg98} suggested that these elevated levels of 
water will persist long after the gas has cooled and any
enhancement of the water abundance may be present over the entire lifetime 
of these outflows.  The relative abundance of \h2o\ is thus a good tracer of 
the dynamical history of the outflowing molecular gas.  Our analysis of the SWAS data  
rules out the possibility that most of the outflow gas has passed through a shock in 
excess of 10 km s$^{-1}$.

The ISO and high velocity SWAS results for some outflows do indicate  
highly elevated water abundances in a small fraction of the outflowing gas mass
that is consistent with moderate velocity  C-shocks.
The current observations however do not have sufficient spatial 
and/or velocity resolution to determine exactly how these emissions might be related.  
It is suggestive that since 
the mass associated with the ISO and high velocity SWAS \h2o\ emission are very
similar that these may arise from the same C-shock.
For these outflows it would be extremely interesting to determine
the spatial and kinematic relation between the ISO emission, the high velocity
SWAS emission and well-established tracers of C-shocks, such as the 
2.12 $\mu$m H$_2$ emission seen in many of these outflows such as 
L1448-mm \citep{davi95} and NGC1333-SVS13 \citep{moli00}.   Future observations
with {\it Herschel} may help out in this regards.

Mechanisms for suppressing the \h2o\ abundance in the postshock gas were explored
by \citet{snel05}.  The only mechanism that is probably relevant for these outflows
is the freeze-out of gas-phase water onto grain mantles.  
This was the primary mechanism in the model of \citet{berg98} for the
reduction of gas-phase water in post-shock gas.  The timescale for significant depletion
of gas-phase water depends on the post shock density of the gas and the total
grain cross-section.  In the model of \citet{berg98}, for gas denser than
10$^6$ cm$^{-3}$, the timescale is shorter than 10$^4$ years, however at
densities of 10$^4$ cm$^{-3}$ the time scale is longer than 3$\times10^5$ 
years.  However since outflow lifetimes are extremely poorly understood, it is 
possible that the freeze-out of water is an important factor.  Based on the 
timescale for water freeze-out, \citet{snel05} provided a simple expression for
the hydrogen column density in the post-shock gas to the point 
where water will be frozen onto grain 
mantles. This calculation uses a slightly larger grain cross-section
than the numerical model of  \citet{berg98}.  Assuming a shock velocity of 10 km s$^{-1}$ and
a shock compression factor of 10, the molecular hydrogen column would be 
of order 10$^{21}$ cm$^{-2}$.  If the relative abundance of ortho-\h2o\ in this
column was as  $>$10$^{-4}$, as predicted by the C-shock models, the 
emission from this gas would greatly
exceed what SWAS observed.  Unless a more efficient mechanism is found for
depleting the gas-phase water in the post-shock gas, we believe that
the \h2o\ abundance in outflows provides strong evidence that the bulk of the
outflowing molecular gas was never subjected to shocks greater than 
10 km s$^{-1}$ that would convert all of gas phase oxygen to water.  
This result places severe restrictions on the mechanism by which the molecular gas is 
accelerated in these outflows and favor mechanisms, such as turbulent 
entrainment, that can accelerate the gas gently.  However, some 
enhancement of the water abundance 
in the bulk of the outflowing gas over that measured in quiescent cloud gas may be
necessary.  Slower shocks ($<$ 10 km s$^{-1}$) could heat the gas and 
dust to temperatures above
the water-ice evaporation temperature, but below the temperature in which
the rapid neutral-neutral reactions are activited, a suggestion made 
by \citet{cern06} for the water abundance measured in Orion KL..

Finally we note that strong [OI] emission at a wavelength of 63 $\mu$m  was 
observed by ISO in all of the outflows.  Many of the papers presenting 
ISO outflow results suggest that the origin of the [OI] emission is in
J-shocks that dissociate the molecular gas.  Alternatively, the [OI] emission could
arise from the same weak shocks that accelerate the bulk of the molecular gas.
Future observations with {\it Herschel}, which has better angular and spectral
resolution, may help determine the relationship between the
\h2o\ and [OI] emissions and other shock tracers in these outflows and provide a
better understanding of the evolution of the \h2o\ abundance in these outflows.

We acknowledge the support by NASA contract NAS5-30702 that funded
SWAS and NSF grant AST 01-00793 that funds the Five College Radio Astronomy Observatory. 
J.F acknowledges support by the NASA Goddard Center for Astrobiology through Cooporative 
Agreement NNG04G155A.  R.L.S., M.J.K., G.J.M., D.J.H, and E.A.B. acknowledge support 
of grant NNG06GB30G from NASA's LTSA prgrams.  D.A.N. acknowledge the support of grant NA65-13114 from
NASA's LTSA program.

\clearpage

\clearpage

\begin{deluxetable}{lcrcc}
\tabletypesize{\footnotesize}
\tablewidth{0pt}
\tablecaption{Positions of Observed Outflows \& SWAS Observing Times}
\tablehead{
\colhead{Source}  & \colhead{$\alpha$(2000)\tablenotemark{1}} & \colhead{$\delta$(2000)\tablenotemark{1}}
& \colhead{Distance} & \colhead{t$_{int}$}\\
\colhead{}  & \colhead{}  & \colhead{}  & \colhead{pc}  & \colhead{hr}}
\startdata
\hline
L1448-mm       & 03 25 30.5 & +30 45 43 & 300  & 31.47 \\
GL490          & 03 27 38.5 & +58 46 58 & 900  & 89.50 \\
NGC1333-SVS13  & 03 29 03.7 & +31 16 03 & 220  & 29.62 \\
Orion KL       & 05 35 14.5 & -05 22 37 & 500  & 8.50  \\
OMC2           & 05 35 27.3 & -05 09 49 & 450  & 6.34  \\
HH25mm         & 05 46 07.3 & -00 13 40 & 400  & 48.02 \\
NGC2071        & 05 47 04.1 & +00 21 43 & 390  & 19.32 \\
Mon R2	       & 06 07 46.7 & -06 22 42 & 950  & 17.55 \\
NGC2264 D      & 06 41 03.9 & +09 34 39 & 800  & 35.55 \\
NGC2264 C      & 06 41 10.7 & +09 29 07 & 800  & 11.36 \\
$\rho$ Oph A   & 16 26 23.4 & -24 23 02 & 160  & 59.68 \\
L1689N         & 16 32 22.7 & -24 28 33 & 120  & 24.12 \\
Ser SMM1       & 18 29 49.6 & +01 15 20 & 310  & 35.60 \\
L1157          & 20 39 06.5 & +68 02 14 & 440  & 55.45 \\
L1228          & 20 57 13.0 & +77 35 47 & 300  & 33.87 \\
IC 1396N       & 21 40 42.3 & +58 16 10 & 750  & 59.02 \\
S140           & 22 19 17.1 & +63 18 46 & 910  & 24.78 \\
Cepheus A HW2  & 22 56 17.9 & +62 01 50 & 725  & 17.62 \\
\hline
\enddata
\tablenotetext{1}{Units of right ascension are hours, minutes, and seconds, and units of declination are degrees, arcminutes, and arcseconds.}
\end{deluxetable}

\begin{deluxetable}{lll@{ $\to$ }rccccl}
\tablewidth{0pt}
\tablecaption{SWAS Line Fluxes and H$_2$O Abundances\tablenotemark{a}}
\tabletypesize{\scriptsize}
\tablehead{
\colhead{Source} & \colhead{} & \multicolumn{2}{c}{$V_{LSR}$ Interval} & \colhead{$\int T_A^*$(H$_2$O)d$v$} & \colhead{H$_2$O Line Flux} & \colhead{$^{12}$CO/$^{13}$CO} & \colhead{Mass} & \colhead{o-H$_2$O Abundance\tablenotemark{b}}\\
\colhead{} & \colhead{} & \multicolumn{2}{c}{(km s$^{-1}$)} & \colhead{(K km s$^{-1}$)} & \colhead{(10$^{-20}$ W cm$^{-2}$)} & \colhead{} & \colhead{(M$_{\odot}$)} & \colhead{}}

\startdata
 L1448-mm  & Blue &  -31.0 &   -1.0 &   0.28 (0.05) &  0.75 (0.13) &   21.4 ( 4.0) & 4.9$\times10^{-1}$ & \phantom{$<$}1.5$\times10^{-6}$\\
           & Red  &    8.0 &   23.0 &   0.27 (0.03) &  0.72 (0.09) &   21.6 ( 4.0) & 2.9$\times10^{-1}$ & \phantom{$<$}3.7$\times10^{-6}$\\[10pt]
     GL490 & Blue &  -42.0 &  -17.0 &   0.57 (0.03) &  1.51 (0.09) &   19.6 ( 0.7) & 1.4$\times10^{1}$ & \phantom{$<$}1.3$\times10^{-7}$\\
           & Red  &   -8.0 &   22.0 &   0.34 (0.04) &  0.90 (0.10) &   27.8 ( 1.7) & 1.2$\times10^{1}$ & \phantom{$<$}9.6$\times10^{-8}$\\[10pt]
NGC1333-SVS13 & Blue &  -20.5 &    4.5 &   0.54 (0.04) &  1.42 (0.11) &   48.3 ( 9.3) & 5.4$\times10^{-1}$ & \phantom{$<$}1.6$\times10^{-6}$\\
           & Red  &   11.5 &   31.5 &   0.64 (0.04) &  1.68 (0.10) &   33.4 ( 3.1) & 6.7$\times10^{-1}$ & \phantom{$<$}1.3$\times10^{-6}$\\[10pt]
  Orion KL & Blue &  -35.5 &    4.5 &   40.5 (0.09) & 106.43 (0.25) &   27.6 ( 0.8) & 1.2$\times10^{1}$ & \phantom{$<$}9.3$\times10^{-6}$\\
           & Red  &   13.5 &   53.5 &   48.1 (0.09) & 126.41 (0.25) &   27.9 ( 0.9) & 9.8 & \phantom{$<$}1.6$\times10^{-5}$\\[10pt]
     OMC-2 & Blue &   -2.5 &    7.5 &   0.84 (0.07) &  2.21 (0.17) &   28.0 ( 3.5) & 3.4 & \phantom{$<$}3.6$\times10^{-7}$\\
           & Red  &   13.5 &   23.5 &   0.82 (0.07) &  2.17 (0.17) &   62.1 (17.7) & 1.1 & \phantom{$<$}8.7$\times10^{-7}$\\[10pt]
    HH25mm & Blue &   -6.5 &    8.5 &   0.36 (0.03) &  0.96 (0.09) &   21.0 ( 0.9) & 4.4 & \phantom{$<$}4.2$\times10^{-7}$\\
           & Red  &   12.0 &   37.0 &   0.50 (0.04) &  1.33 (0.12) &   25.2 ( 1.6) & 4.5 & \phantom{$<$}7.0$\times10^{-7}$\\[10pt]
   NGC2071 & Blue &  -23.5 &    6.5 &   1.04 (0.07) &  2.73 (0.17) &   25.7 ( 0.8) & 6.8 & \phantom{$<$}1.1$\times10^{-7}$\\
           & Red  &   12.5 &   37.5 &   1.73 (0.06) &  4.55 (0.16) &   33.4 ( 1.2) & 6.9 & \phantom{$<$}2.1$\times10^{-7}$\\[10pt]
     MonR2 & Blue &   -2.2 &    7.8 &   0.47 (0.04) &  1.23 (0.10) &   10.3 ( 0.1) & 1.2$\times10^{2}$ & \phantom{$<$}1.8$\times10^{-8}$\\
           & Red  &   13.0 &   28.0 &   0.59 (0.05) &  1.56 (0.12) &    8.6 ( 0.1) & 2.6$\times10^{2}$ & \phantom{$<$}1.3$\times10^{-8}$\\[10pt]
 NGC2264 D & Blue &   -9.0 &    1.0 &   0.13 (0.03) &  0.33 (0.07) &  \nodata      & 1.8 & \phantom{$<$}1.4$\times10^{-6}$\\
           & Red  &   10.0 &   20.0 &   0.19 (0.03) &  0.51 (0.07) &  \nodata      & 1.9 & \phantom{$<$}2.4$\times10^{-6}$\\[10pt]
 NGC2264 C & Red  &   11.0 &   21.0 &   0.83 (0.04) &  2.18 (0.11) &   21.1 ( 1.1) & 7.2 & \phantom{$<$}1.7$\times10^{-6}$\\[10pt]
$\rho$ Oph A & Blue &   -3.5 &    1.5 &   0.28 (0.01) &  0.74 (0.04) &   17.8 ( 2.7) & 2.2$\times10^{-1}$ & \phantom{$<$}9.4$\times10^{-7}$\\
           & Red  &    6.0 &   11.0 &   0.17 (0.01) &  0.43 (0.04) &   13.3 ( 1.1) & 4.3$\times10^{-1}$ & \phantom{$<$}3.7$\times10^{-7}$\\[10pt]
    L1689N & Blue &  -12.5 &    2.5 &   0.57 (0.05) &  1.49 (0.12) &   18.8 ( 2.4) & 2.7$\times10^{-1}$ & \phantom{$<$}9.4$\times10^{-7}$\\
           & Red  &    6.0 &   21.0 &   1.08 (0.05) &  2.84 (0.12) &   20.8 ( 3.0) & 1.1$\times10^{-1}$ & \phantom{$<$}2.9$\times10^{-6}$\\[10pt]
  Ser SMM1 & Blue &   -4.5 &    5.5 &   0.40 (0.02) &  1.05 (0.06) &   19.8 ( 1.3) & 1.8 & \phantom{$<$}7.1$\times10^{-7}$\\
           & Red  &   11.0 &   21.0 &   0.32 (0.02) &  0.83 (0.06) &   20.0 ( 1.1) & 2.9 & \phantom{$<$}3.8$\times10^{-7}$\\[10pt]
     L1157 & Blue &  -13.5 &    1.5 &   0.70 (0.02) &  1.84 (0.06) &   28.7 ( 5.1) & 4.1$\times10^{-1}$ & \phantom{$<$}8.0$\times10^{-6}$\\
           & Red  &    4.0 &   29.0 &   0.60 (0.03) &  1.58 (0.08) &  \nodata      & 3.2$\times10^{-1}$ & \phantom{$<$}9.7$\times10^{-6}$\\[10pt]
     L1228 & Blue &  -19.5 &   -9.5 &   0.17 (0.02) &  0.45 (0.06) &   21.7 ( 1.3) & 8.7$\times10^{-1}$ & \phantom{$<$}5.9$\times10^{-7}$\\
           & Red  &   -6.0 &    4.0 &   0.14 (0.02) &  0.38 (0.06) &   20.7 ( 2.5) & 5.1$\times10^{-1}$ & \phantom{$<$}8.5$\times10^{-7}$\\[10pt]
   IC1396N & Blue &  -18.0 &   -3.0 &   0.36 (0.02) &  0.95 (0.06) &   31.3 ( 6.0) & 1.5 & \phantom{$<$}2.7$\times10^{-6}$\\
           & Red  &    4.0 &   19.0 &   0.33 (0.02) &  0.88 (0.06) &   36.2 ( 9.1) & 1.2 & \phantom{$<$}3.2$\times10^{-6}$\\[10pt]
      S140 & Blue &  -36.0 &  -11.0 &   0.24 (0.04) &  0.63 (0.11) &   15.7 ( 0.4) & 3.2$\times10^{1}$ & \phantom{$<$}2.4$\times10^{-8}$\\
           & Red  &   -4.0 &    6.0 &   0.47 (0.03) &  1.25 (0.07) &   18.3 ( 0.6) & 1.6$\times10^{1}$ & \phantom{$<$}1.0$\times10^{-7}$\\[10pt]
Ceph A HW2 & Blue &  -35.0 &  -15.0 &   1.34 (0.06) &  3.53 (0.16) &   14.8 ( 0.5) & 1.8$\times10^{1}$ & \phantom{$<$}1.9$\times10^{-7}$\\
           & Red  &   -6.0 &   14.0 &   1.27 (0.06) &  3.35 (0.16) &   19.0 ( 0.9) & 1.7$\times10^{1}$ & \phantom{$<$}2.0$\times10^{-7}$\\[10pt]

\enddata
\tablenotetext{a} {The numbers in parantheses are 1-$\sigma$ statistical uncertainties on
the integrated intensity, line flux and observed isotopic ratio}
\tablenotetext{b}{o-H$_2$O abundances relative to H$_2$ assuming a H$_2$/CO ratio of 10$^4$}
\end{deluxetable}

\begin{deluxetable}{lll@{ $\to$ }rccccl}
\tablewidth{0pt}
\tablecaption{SWAS Line Fluxes and H$_2$O Abundances as a Function of Velocity\tablenotemark{a}}
\tabletypesize{\scriptsize}
\tablehead{
\colhead{Source} & \colhead{} & \multicolumn{2}{c}{$V_{LSR}$ Interval} & \colhead{$\int T_A^*$(H$_2$O)d$v$} & \colhead{H$_2$O Line Flux} & \colhead{$^{12}$CO/$^{13}$CO} & \colhead{Mass} & \colhead{o-H$_2$O Abundance\tablenotemark{b}}\\
\colhead{} & \colhead{} & \multicolumn{2}{c}{(km s$^{-1}$)} & \colhead{(K km s$^{-1}$)} & \colhead{(10$^{-20}$ W cm$^{-2}$)} & \colhead{} & \colhead{(M$_{\odot}$)} & \colhead{}}

\startdata
 L1448-mm & Blue &   -6.0 &   -1.0 &   0.07 (0.02) &  0.18 (0.05) &   31.3 ( 9.3) & 1.6$\times10^{-1}$ & \phantom{$<$}1.2$\times10^{-6}$\\
           &      &  -11.0 &   -6.0 &   $<$  0.06   &   $<$ 0.16   &  \nodata      & 2.1$\times10^{-2}$ & $<$7.1$\times10^{-6}$\\
           &      &  -16.0 &  -11.0 &   0.06 (0.02) &  0.16 (0.05) &  \nodata      & 9.7$\times10^{-3}$ & \phantom{$<$}1.2$\times10^{-5}$\\
           &      &  -21.0 &  -16.0 &   0.06 (0.02) &  0.16 (0.05) &  \nodata      & 9.4$\times10^{-3}$ & \phantom{$<$}1.3$\times10^{-5}$\\
           &      &  -26.0 &  -21.0 &   $<$  0.06   &   $<$ 0.16   &  \nodata      & 2.6$\times10^{-3}$ & $<$4.5$\times10^{-5}$\\
           &      &  -31.0 &  -26.0 &   $<$  0.06   &   $<$ 0.16   &  \nodata      & 3.2$\times10^{-3}$ & $<$4.4$\times10^{-5}$\\
           & Red  &    8.0 &   13.0 &   0.12 (0.02) &  0.31 (0.05) &   36.8 ( 9.0) & 1.3$\times10^{-1}$ & \phantom{$<$}3.5$\times10^{-6}$\\
           &      &   13.0 &   18.0 &   0.08 (0.02) &  0.21 (0.05) &    9.0 ( 2.4) & 8.3$\times10^{-2}$ & \phantom{$<$}4.3$\times10^{-6}$\\
           &      &   18.0 &   23.0 &   0.08 (0.02) &  0.20 (0.05) &  \nodata      & 4.3$\times10^{-3}$ & \phantom{$<$}1.1$\times10^{-4}$\\[10pt]
     GL490 & Blue &  -22.0 &  -17.0 &   0.15 (0.01) &  0.39 (0.04) &   19.1 ( 0.5) & 9.9 & \phantom{$<$}4.8$\times10^{-8}$\\
           &      &  -27.0 &  -22.0 &   0.11 (0.01) &  0.29 (0.04) &   21.2 ( 1.5) & 3.1 & \phantom{$<$}1.1$\times10^{-7}$\\
           &      &  -32.0 &  -27.0 &   0.13 (0.01) &  0.34 (0.04) &   28.2 ( 6.7) & 6.7$\times10^{-1}$ & \phantom{$<$}5.4$\times10^{-7}$\\
           &      &  -37.0 &  -32.0 &   0.08 (0.01) &  0.22 (0.04) &  \nodata      & 1.3$\times10^{-1}$ & \phantom{$<$}1.8$\times10^{-6}$\\
           &      &  -42.0 &  -37.0 &   0.10 (0.01) &  0.28 (0.04) &  \nodata      & 3.4$\times10^{-2}$ & \phantom{$<$}8.3$\times10^{-6}$\\
           & Red  &   -8.0 &   -3.0 &   0.10 (0.01) &  0.26 (0.04) &   24.0 ( 0.9) & 8.8 & \phantom{$<$}3.9$\times10^{-8}$\\
           &      &   -3.0 &    2.0 &   0.08 (0.01) &  0.22 (0.04) &   44.6 ( 7.2) & 1.4 & \phantom{$<$}1.7$\times10^{-7}$\\
           &      &    2.0 &    7.0 &   0.05 (0.01) &  0.14 (0.04) &   34.3 (11.3) & 5.3$\times10^{-1}$ & \phantom{$<$}2.7$\times10^{-7}$\\
           &      &    7.0 &   12.0 &   $<$  0.04   &   $<$ 0.12   &  \nodata      & 1.6$\times10^{-1}$ & $<$7.5$\times10^{-7}$\\
           &      &   12.0 &   17.0 &   0.07 (0.01) &  0.18 (0.04) &  \nodata      & 8.3$\times10^{-2}$ & \phantom{$<$}2.5$\times10^{-6}$\\
           &      &   17.0 &   22.0 &   $<$  0.04   &   $<$ 0.12   &  \nodata      & 2.6$\times10^{-2}$ & $<$4.5$\times10^{-6}$\\[10pt]
NGC1333-SVS13 & Blue &   -0.5 &    4.5 &   0.20 (0.02) &  0.53 (0.05) &   34.9 ( 3.1) & 6.7$\times10^{-1}$ & \phantom{$<$}5.1$\times10^{-7}$\\
           &      &   -5.5 &   -0.5 &   0.13 (0.02) &  0.34 (0.05) &  \nodata      & 2.6$\times10^{-2}$ & \phantom{$<$}5.7$\times10^{-6}$\\
           &      &  -10.5 &   -5.5 &   0.09 (0.02) &  0.24 (0.05) &  \nodata      & 5.5$\times10^{-3}$ & \phantom{$<$}1.6$\times10^{-5}$\\
           &      &  -15.5 &  -10.5 &   0.12 (0.02) &  0.30 (0.05) &  \nodata      & 2.4$\times10^{-3}$ & \phantom{$<$}4.8$\times10^{-5}$\\
           &      &  -20.5 &  -15.5 &   $<$  0.05   &   $<$ 0.14   &  \nodata      & 1.5$\times10^{-3}$ & $<$3.1$\times10^{-5}$\\
           & Red  &   11.5 &   16.5 &   0.30 (0.02) &  0.79 (0.05) &   30.3 ( 1.7) & 5.7$\times10^{-1}$ & \phantom{$<$}7.2$\times10^{-7}$\\
           &      &   16.5 &   21.5 &   0.17 (0.02) &  0.44 (0.05) &  \nodata      & 4.4$\times10^{-2}$ & \phantom{$<$}4.6$\times10^{-6}$\\
           &      &   21.5 &   26.5 &   0.10 (0.02) &  0.27 (0.05) &  \nodata      & 8.7$\times10^{-3}$ & \phantom{$<$}1.5$\times10^{-5}$\\
           &      &   26.5 &   31.5 &   0.07 (0.02) &  0.18 (0.05) &  \nodata      & 1.2$\times10^{-3}$ & \phantom{$<$}7.9$\times10^{-5}$\\[10pt]
  Orion KL & Blue &   -0.5 &    4.5 &   10.2 (0.03) & 26.92 (0.09) &   20.2 ( 0.4) & 7.8 & \phantom{$<$}2.4$\times10^{-6}$\\
           &      &   -5.5 &   -0.5 &   8.28 (0.03) & 21.77 (0.09) &   25.1 ( 1.1) & 2.2 & \phantom{$<$}6.4$\times10^{-6}$\\
           &      &  -10.5 &   -5.5 &   6.30 (0.03) & 16.57 (0.09) &   34.4 ( 3.1) & 1.2 & \phantom{$<$}8.1$\times10^{-6}$\\
           &      &  -15.5 &  -10.5 &   4.98 (0.03) & 13.11 (0.09) &   49.2 ( 9.5) & 5.3$\times10^{-1}$ & \phantom{$<$}1.3$\times10^{-5}$\\
           &      &  -20.5 &  -15.5 &   3.80 (0.03) &  9.99 (0.09) &  \nodata      & 2.6$\times10^{-1}$ & \phantom{$<$}1.9$\times10^{-5}$\\
           &      &  -25.5 &  -20.5 &   3.01 (0.03) &  7.92 (0.09) &  \nodata      & 1.1$\times10^{-1}$ & \phantom{$<$}3.6$\times10^{-5}$\\
           &      &  -30.5 &  -25.5 &   2.26 (0.03) &  5.94 (0.09) &  \nodata      & 3.8$\times10^{-2}$ & \phantom{$<$}9.2$\times10^{-5}$\\
           &      &  -35.5 &  -30.5 &   1.60 (0.03) &  4.21 (0.09) &  \nodata      & 1.8$\times10^{-2}$ & \phantom{$<$}1.2$\times10^{-4}$\\
           & Red  &   13.5 &   18.5 &   13.7 (0.03) & 36.12 (0.09) &   24.1 ( 0.5) & 6.7 & \phantom{$<$}4.3$\times10^{-6}$\\
           &      &   18.5 &   23.5 &   10.8 (0.03) & 28.36 (0.09) &   27.2 ( 1.3) & 1.8 & \phantom{$<$}1.3$\times10^{-5}$\\
           &      &   23.5 &   28.5 &   8.22 (0.03) & 21.61 (0.09) &   28.3 ( 2.4) & 8.9$\times10^{-1}$ & \phantom{$<$}1.8$\times10^{-5}$\\
           &      &   28.5 &   33.5 &   5.94 (0.03) & 15.61 (0.09) &  \nodata      & 2.2$\times10^{-1}$ & \phantom{$<$}4.9$\times10^{-5}$\\
           &      &   33.5 &   38.5 &   3.99 (0.03) & 10.50 (0.09) &  \nodata      & 1.1$\times10^{-1}$ & \phantom{$<$}6.6$\times10^{-5}$\\
           &      &   38.5 &   43.5 &   2.61 (0.03) &  6.87 (0.09) &  \nodata      & 4.5$\times10^{-2}$ & \phantom{$<$}9.6$\times10^{-5}$\\
           &      &   43.5 &   48.5 &   1.70 (0.03) &  4.47 (0.09) &  \nodata      & 2.8$\times10^{-2}$ & \phantom{$<$}8.5$\times10^{-5}$\\
           &      &   48.5 &   53.5 &   1.09 (0.03) &  2.86 (0.09) &  \nodata      & 2.2$\times10^{-2}$ & \phantom{$<$}8.1$\times10^{-5}$\\[10pt]
      OMC2 & Blue &    2.5 &    7.5 &   0.46 (0.05) &  1.22 (0.12) &   21.7 ( 1.6) & 4.2 & \phantom{$<$}1.6$\times10^{-7}$\\
           &      &   -2.5 &    2.5 &   0.38 (0.05) &  0.99 (0.12) &  \nodata      & 1.9$\times10^{-2}$ & \phantom{$<$}1.7$\times10^{-5}$\\
           & Red  &   13.5 &   18.5 &   0.56 (0.05) &  1.46 (0.12) &   52.4 ( 9.7) & 1.3 & \phantom{$<$}5.3$\times10^{-7}$\\
           &      &   18.5 &   23.5 &   0.27 (0.05) &  0.71 (0.12) &  \nodata      & 1.6$\times10^{-2}$ & \phantom{$<$}1.3$\times10^{-5}$\\[10pt]
    HH25mm & Blue &    3.5 &    8.5 &   0.16 (0.02) &  0.42 (0.05) &   17.8 ( 0.4) & 4.8 & \phantom{$<$}1.7$\times10^{-7}$\\
           &      &   -1.5 &    3.5 &   0.12 (0.02) &  0.33 (0.05) &  \nodata      & 7.6$\times10^{-2}$ & \phantom{$<$}7.0$\times10^{-6}$\\
           &      &   -6.5 &   -1.5 &   0.08 (0.02) &  0.21 (0.05) &  \nodata      & 7.8$\times10^{-3}$ & \phantom{$<$}3.0$\times10^{-5}$\\
           & Red  &   12.0 &   17.0 &   0.19 (0.02) &  0.49 (0.05) &   26.4 ( 1.1) & 3.6 & \phantom{$<$}3.0$\times10^{-7}$\\
           &      &   17.0 &   22.0 &   0.16 (0.02) &  0.42 (0.05) &   37.3 ( 8.2) & 3.1$\times10^{-1}$ & \phantom{$<$}3.8$\times10^{-6}$\\
           &      &   22.0 &   27.0 &   0.09 (0.02) &  0.23 (0.05) &  \nodata      & 3.4$\times10^{-2}$ & \phantom{$<$}2.7$\times10^{-5}$\\
           &      &   27.0 &   32.0 &   $<$  0.06   &   $<$ 0.16   &  \nodata      & 8.2$\times10^{-3}$ & $<$1.0$\times10^{-4}$\\
           &      &   32.0 &   37.0 &   $<$  0.06   &   $<$ 0.16   &  \nodata      & 1.3$\times10^{-3}$ & $<$7.2$\times10^{-4}$\\[10pt]
   NGC2071 & Blue &    1.5 &    6.5 &   0.35 (0.03) &  0.91 (0.07) &   20.9 ( 0.4) & 6.4 & \phantom{$<$}4.2$\times10^{-8}$\\
           &      &   -3.5 &    1.5 &   0.29 (0.03) &  0.75 (0.07) &   31.7 ( 2.0) & 8.9$\times10^{-1}$ & \phantom{$<$}2.1$\times10^{-7}$\\
           &      &   -8.5 &   -3.5 &   0.17 (0.03) &  0.43 (0.07) &  \nodata      & 9.0$\times10^{-2}$ & \phantom{$<$}1.1$\times10^{-6}$\\
           &      &  -13.5 &   -8.5 &   $<$  0.08   &   $<$ 0.21   &  \nodata      & 2.6$\times10^{-2}$ & $<$1.7$\times10^{-6}$\\
           &      &  -18.5 &  -13.5 &   0.16 (0.03) &  0.42 (0.07) &  \nodata      & 1.6$\times10^{-2}$ & \phantom{$<$}5.5$\times10^{-6}$\\
           &      &  -23.5 &  -18.5 &   $<$  0.08   &   $<$ 0.21   &  \nodata      & 3.2$\times10^{-3}$ & $<$1.4$\times10^{-5}$\\
           & Red  &   12.5 &   17.5 &   0.81 (0.03) &  2.14 (0.07) &   26.4 ( 0.4) & 7.2 & \phantom{$<$}9.9$\times10^{-8}$\\
           &      &   17.5 &   22.5 &   0.41 (0.03) &  1.09 (0.07) &   60.1 ( 9.1) & 4.7$\times10^{-1}$ & \phantom{$<$}6.1$\times10^{-7}$\\
           &      &   22.5 &   27.5 &   0.25 (0.03) &  0.65 (0.07) &  \nodata      & 8.8$\times10^{-2}$ & \phantom{$<$}1.8$\times10^{-6}$\\
           &      &   27.5 &   32.5 &   0.16 (0.03) &  0.42 (0.07) &  \nodata      & 3.2$\times10^{-2}$ & \phantom{$<$}3.0$\times10^{-6}$\\
           &      &   32.5 &   37.5 &   0.10 (0.03) &  0.26 (0.07) &  \nodata      & 8.5$\times10^{-3}$ & \phantom{$<$}6.6$\times10^{-6}$\\[10pt]
     MonR2 & Blue &    2.8 &    7.8 &   0.38 (0.03) &  1.01 (0.07) &    9.9 ( 0.1) & 1.3$\times10^{2}$ & \phantom{$<$}1.5$\times10^{-8}$\\
           &      &   -2.2 &    2.8 &   0.08 (0.03) &  0.22 (0.07) &  \nodata      & 1.4$\times10^{-1}$ & \phantom{$<$}2.7$\times10^{-6}$\\
           & Red  &   13.0 &   18.0 &   0.45 (0.03) &  1.19 (0.07) &    8.0 ( 0.0) & 2.4$\times10^{2}$ & \phantom{$<$}1.1$\times10^{-8}$\\
           &      &   18.0 &   23.0 &   0.09 (0.03) &  0.24 (0.07) &   17.9 ( 1.7) & 1.3$\times10^{1}$ & \phantom{$<$}3.1$\times10^{-8}$\\
           &      &   23.0 &   28.0 &   $<$  0.08   &   $<$ 0.21   &  \nodata      & 2.4$\times10^{-1}$ & $<$1.2$\times10^{-6}$\\[10pt]
 NGC2264 D & Blue &   -4.0 &    1.0 &   0.09 (0.02) &  0.24 (0.05) &  \nodata      & 1.6 & \phantom{$<$}1.1$\times10^{-6}$\\
           &      &   -9.0 &   -4.0 &   $<$  0.05   &   $<$ 0.14   &  \nodata      & 5.2$\times10^{-2}$ & $<$1.6$\times10^{-5}$\\
           & Red  &   10.0 &   15.0 &   0.13 (0.02) &  0.35 (0.05) &  \nodata      & 1.8 & \phantom{$<$}1.6$\times10^{-6}$\\
           &      &   15.0 &   20.0 &   0.06 (0.02) &  0.16 (0.05) &  \nodata      & 7.5$\times10^{-2}$ & \phantom{$<$}2.9$\times10^{-5}$\\[10pt]
 NGC2264 C & Red  &   11.0 &   16.0 &   0.52 (0.03) &  1.37 (0.08) &   22.1 ( 1.0) & 6.3 & \phantom{$<$}1.2$\times10^{-6}$\\
           &      &   16.0 &   21.0 &   0.31 (0.03) &  0.80 (0.08) &   14.9 ( 4.1) & 5.8$\times10^{-1}$ & \phantom{$<$}6.4$\times10^{-6}$\\[10pt]
$\rho$ Oph A & Blue &   -3.5 &    1.5 &   0.28 (0.01) &  0.74 (0.04) &   17.8 ( 2.7) & 2.2$\times10^{-1}$ & \phantom{$<$}9.5$\times10^{-7}$\\
           & Red  &    6.0 &   11.0 &   0.17 (0.01) &  0.43 (0.04) &   13.3 ( 1.1) & 4.3$\times10^{-1}$ & \phantom{$<$}3.7$\times10^{-7}$\\[10pt]
    L1689N & Blue &   -2.5 &    2.5 &   0.37 (0.03) &  0.97 (0.07) &   18.3 ( 1.8) & 2.2$\times10^{-1}$ & \phantom{$<$}7.5$\times10^{-7}$\\
           &      &   -7.5 &   -2.5 &   0.13 (0.03) &  0.33 (0.07) &  \nodata      & 6.8$\times10^{-3}$ & \phantom{$<$}8.4$\times10^{-6}$\\
           &      &  -12.5 &   -7.5 &   $<$  0.08   &   $<$ 0.21   &  \nodata      & 9.9$\times10^{-4}$ & $<$3.3$\times10^{-5}$\\
           & Red  &    6.0 &   11.0 &   0.72 (0.03) &  1.89 (0.07) &   21.3 ( 2.5) & 9.0$\times10^{-2}$ & \phantom{$<$}2.5$\times10^{-6}$\\
           &      &   11.0 &   16.0 &   0.25 (0.03) &  0.66 (0.07) &  \nodata      & 4.7$\times10^{-3}$ & \phantom{$<$}1.6$\times10^{-5}$\\
           &      &   16.0 &   21.0 &   0.11 (0.03) &  0.29 (0.07) &  \nodata      & 4.9$\times10^{-4}$ & \phantom{$<$}8.1$\times10^{-5}$\\[10pt]
  Ser SMM1 & Blue &    0.5 &    5.5 &   0.27 (0.01) &  0.70 (0.04) &   17.8 ( 0.8) & 1.9 & \phantom{$<$}4.7$\times10^{-7}$\\
           &      &   -4.5 &    0.5 &   0.13 (0.01) &  0.35 (0.04) &  \nodata      & 1.9$\times10^{-2}$ & \phantom{$<$}1.6$\times10^{-5}$\\
           & Red  &   11.0 &   16.0 &   0.18 (0.01) &  0.48 (0.04) &   17.9 ( 0.6) & 3.0 & \phantom{$<$}2.1$\times10^{-7}$\\
           &      &   16.0 &   21.0 &   0.14 (0.01) &  0.36 (0.04) &  \nodata      & 3.6$\times10^{-2}$ & \phantom{$<$}1.4$\times10^{-5}$\\[10pt]
     L1157 & Blue &   -3.5 &    1.5 &   0.33 (0.01) &  0.86 (0.04) &   35.4 ( 5.6) & 2.8$\times10^{-1}$ & \phantom{$<$}5.3$\times10^{-6}$\\
           &      &   -8.5 &   -3.5 &   0.22 (0.01) &  0.58 (0.04) &  \nodata      & 1.9$\times10^{-2}$ & \phantom{$<$}4.9$\times10^{-5}$\\
           &      &  -13.5 &   -8.5 &   0.15 (0.01) &  0.40 (0.04) &  \nodata      & 3.9$\times10^{-3}$ & \phantom{$<$}3.5$\times10^{-4}$\\
           & Red  &    4.0 &    9.0 &   0.23 (0.01) &  0.59 (0.04) &   33.7 ( 7.0) & 2.9$\times10^{-1}$ & \phantom{$<$}3.8$\times10^{-6}$\\
           &      &    9.0 &   14.0 &   0.14 (0.01) &  0.37 (0.04) &  \nodata      & 5.2$\times10^{-2}$ & \phantom{$<$}1.3$\times10^{-5}$\\
           &      &   14.0 &   19.0 &   0.11 (0.01) &  0.28 (0.04) &  \nodata      & 3.6$\times10^{-2}$ & \phantom{$<$}1.5$\times10^{-5}$\\
           &      &   19.0 &   24.0 &   0.06 (0.01) &  0.16 (0.04) &  \nodata      & 2.0$\times10^{-2}$ & \phantom{$<$}1.6$\times10^{-5}$\\
           &      &   24.0 &   29.0 &   0.07 (0.01) &  0.18 (0.04) &  \nodata      & 7.0$\times10^{-3}$ & \phantom{$<$}5.2$\times10^{-5}$\\[10pt]
     L1228 & Blue &  -14.5 &   -9.5 &   0.09 (0.01) &  0.24 (0.04) &   20.9 ( 1.0) & 8.4$\times10^{-1}$ & \phantom{$<$}3.2$\times10^{-7}$\\
           &      &  -19.5 &  -14.5 &   0.08 (0.01) &  0.22 (0.04) &  \nodata      & 2.2$\times10^{-2}$ & \phantom{$<$}1.2$\times10^{-5}$\\
           & Red  &   -6.0 &   -1.0 &   0.08 (0.01) &  0.21 (0.04) &   21.1 ( 2.0) & 5.1$\times10^{-1}$ & \phantom{$<$}4.5$\times10^{-7}$\\
           &      &   -1.0 &    4.0 &   0.07 (0.01) &  0.17 (0.04) &  \nodata      & 5.3$\times10^{-3}$ & \phantom{$<$}4.4$\times10^{-5}$\\[10pt]
   IC1396N & Blue &   -8.0 &   -3.0 &   0.17 (0.01) &  0.46 (0.04) &   30.9 ( 4.4) & 1.2 & \phantom{$<$}1.6$\times10^{-6}$\\
           &      &  -13.0 &   -8.0 &   0.10 (0.01) &  0.27 (0.04) &  \nodata      & 1.1$\times10^{-1}$ & \phantom{$<$}1.2$\times10^{-5}$\\
           &      &  -18.0 &  -13.0 &   0.08 (0.01) &  0.22 (0.04) &  \nodata      & 1.6$\times10^{-2}$ & \phantom{$<$}5.8$\times10^{-5}$\\
           & Red  &    4.0 &    9.0 &   0.20 (0.01) &  0.52 (0.04) &   30.3 ( 4.4) & 1.4 & \phantom{$<$}1.7$\times10^{-6}$\\
           &      &    9.0 &   14.0 &   0.10 (0.01) &  0.26 (0.04) &  \nodata      & 3.6$\times10^{-2}$ & \phantom{$<$}2.6$\times10^{-5}$\\
           &      &   14.0 &   19.0 &   $<$  0.04   &   $<$ 0.11   &  \nodata      & 4.5$\times10^{-3}$ & $<$8.5$\times10^{-5}$\\[10pt]
      S140 & Blue &  -16.0 &  -11.0 &   0.14 (0.02) &  0.37 (0.05) &   13.8 ( 0.2) & 3.0$\times10^{1}$ & \phantom{$<$}1.5$\times10^{-8}$\\
           &      &  -21.0 &  -16.0 &   $<$  0.05   &   $<$ 0.14   &   21.0 ( 1.3) & 3.8 & $<$4.1$\times10^{-8}$\\
           &      &  -26.0 &  -21.0 &   $<$  0.05   &   $<$ 0.14   &  \nodata      & 1.6$\times10^{-1}$ & $<$8.7$\times10^{-7}$\\
           &      &  -31.0 &  -26.0 &   0.09 (0.02) &  0.25 (0.05) &  \nodata      & 2.8$\times10^{-2}$ & \phantom{$<$}9.0$\times10^{-6}$\\
           &      &  -36.0 &  -31.0 &   $<$  0.05   &   $<$ 0.14   &  \nodata      & 1.5$\times10^{-2}$ & $<$1.5$\times10^{-5}$\\
           & Red  &   -4.0 &    1.0 &   0.42 (0.02) &  1.09 (0.05) &   17.3 ( 0.5) & 1.5$\times10^{1}$ & \phantom{$<$}9.7$\times10^{-8}$\\
           &      &    1.0 &    6.0 &   0.06 (0.02) &  0.15 (0.05) &  \nodata      & 4.7$\times10^{-1}$ & \phantom{$<$}4.9$\times10^{-7}$\\[10pt]
 Ceph A HW2 & Blue &  -20.0 &  -15.0 &   0.62 (0.03) &  1.63 (0.08) &   12.0 ( 0.2) & 1.9$\times10^{1}$ & \phantom{$<$}8.5$\times10^{-8}$\\
           &      &  -25.0 &  -20.0 &   0.34 (0.03) &  0.90 (0.08) &   29.0 ( 5.3) & 8.8$\times10^{-1}$ & \phantom{$<$}8.5$\times10^{-7}$\\
           &      &  -30.0 &  -25.0 &   0.25 (0.03) &  0.65 (0.08) &  \nodata      & 1.1$\times10^{-1}$ & \phantom{$<$}4.7$\times10^{-6}$\\
           &      &  -35.0 &  -30.0 &   0.13 (0.03) &  0.35 (0.08) &  \nodata      & 4.6$\times10^{-2}$ & \phantom{$<$}5.5$\times10^{-6}$\\
           & Red  &   -6.0 &   -1.0 &   0.74 (0.03) &  1.94 (0.08) &   15.3 ( 0.4) & 1.9$\times10^{1}$ & \phantom{$<$}1.1$\times10^{-7}$\\
           &      &   -1.0 &    4.0 &   0.40 (0.03) &  1.04 (0.08) &  \nodata      & 4.4$\times10^{-1}$ & \phantom{$<$}2.1$\times10^{-6}$\\
           &      &    4.0 &    9.0 &   $<$  0.09   &   $<$ 0.24   &  \nodata      & 3.1$\times10^{-1}$ & $<$8.6$\times10^{-7}$\\
           &      &    9.0 &   14.0 &   $<$  0.09   &   $<$ 0.24   &  \nodata      & 2.8$\times10^{-2}$ & $<$6.4$\times10^{-6}$\\[10pt]

\enddata
\tablenotetext{a} {The numbers in parantheses are 1-$\sigma$ statistical uncertainties on
the integrated intensity, line flux and observed isotopic ratio.  All limits quoted are 3-$\sigma$}
\tablenotetext{b}{o-H$_2$O abundances relative to H$_2$ assuming a H$_2$/CO ratio of 10$^4$}
\end{deluxetable}

\begin{deluxetable}{lcccc}
\tabletypesize{\footnotesize}
\tablewidth{0pt}
\tablecaption{ISO Results}
\tablehead{
\colhead{Source} & \colhead{Flux(o-H$_2$O 179 $\mu$m)} & \colhead{Solid Angle} & \colhead{Mass}
& \colhead{H$_2$O Flux Ratio (179 $\mu$m/538 $\mu$m)}\\
\colhead{}  &   \colhead{10$^{-20}$ W cm$^{-2}$} & \colhead{sr}  & \colhead{M$_{\sun}$}}
\startdata
L1448-mm	       &    26.9$\pm$3.4\tablenotemark{a}  &  4.3$\times10^{-10}$ & 4$\times10^{-3}$  &  18.3\\
NGC1333-SVS13  &    14.7$\pm$4.1\tablenotemark{b}  &  1.8$\times10^{-10}$ & 8$\times10^{-4}$  &  4.7\\
HH25mm	       &    10$\pm$2      &  1.2$\times10^{-10}$ & 2$\times10^{-3}$  &  4.4\\
L1689N	       &    21$\pm$2      &  2.6$\times10^{-10}$ & 3$\times10^{-4}$  &  4.8\\
Ser SMM1       &    14.0$\pm$3.2  &  1.7$\times10^{-10}$ & 2$\times10^{-3}$  &  7.4\\
L1157	       &    26.5$\pm$3.5\tablenotemark{c}  &  2.7$\times10^{-10}$ & 5$\times10^{-3}$  &  7.7\\
Orion KL       &   2550$\pm$300   &  3.1$\times10^{-8}$  & 5$\times10^{-1}$  &  10.7\\
\enddata
\tablenotetext{a}{Includes positions L1448-mm and IRS3}
\tablenotetext{b}{Includes positions SVS13 and HH7}
\tablenotetext{c}{Includes positions ON and Blue}
\end{deluxetable}

\begin{deluxetable}{llcr}
\tablewidth{0pt}
\tablecaption{SWAS CI Abundances\tablenotemark{a}}
\tabletypesize{\scriptsize}
\tablehead{
\colhead{Source} & \colhead{} & \colhead{$\int T_A^*$([CI])d$v$} & 
\colhead{CI Abundance\tablenotemark{b}}\\
\colhead{} & \colhead{} & \colhead{(K km s$^{-1}$)} & \colhead{}}

\startdata
L1448-mm & Blue &   $<$  0.14   &$<$1.5$\times10^{-5}$\\
& Red  &   $<$  0.10   &$<$2.5$\times10^{-5}$\\[10pt]
GL490 & Blue &   0.49 (0.06) &\phantom{$<$}1.3$\times10^{-5}$\\
& Red  &   0.53 (0.06) &\phantom{$<$}1.7$\times10^{-5}$\\[10pt]
NGC1333-SVS13 & Blue &   0.55 (0.05) &\phantom{$<$}3.2$\times10^{-5}$\\
& Red  &   $<$  0.14   &$<$5.7$\times10^{-6}$\\[10pt]
Orion KL & Blue &   1.16 (0.03) &\phantom{$<$}6.7$\times10^{-6}$\\
& Red  &   2.04 (0.03) &\phantom{$<$}1.4$\times10^{-5}$\\[10pt]
OMC-2 & Blue &   $<$  0.21   &$<$9.7$\times10^{-6}$\\
& Red  &   $<$  0.21   &$<$2.4$\times10^{-5}$\\[10pt]
HH25mm & Blue &   0.44 (0.04) &\phantom{$<$}1.0$\times10^{-5}$\\
& Red  &   0.49 (0.05) &\phantom{$<$}1.3$\times10^{-5}$\\[10pt]
NGC2071 & Blue &   0.49 (0.06) &\phantom{$<$}6.0$\times10^{-6}$\\
& Red  &   0.49 (0.06) &\phantom{$<$}6.5$\times10^{-6}$\\[10pt]
MonR2 & Blue &   2.27 (0.05) &\phantom{$<$}1.0$\times10^{-5}$\\
& Red  &   3.80 (0.07) &\phantom{$<$}8.9$\times10^{-6}$\\[10pt]
NGC2264 D & Blue &   0.26 (0.03) &\phantom{$<$}5.9$\times10^{-5}$\\
& Red  &   0.29 (0.03) &\phantom{$<$}7.2$\times10^{-5}$\\[10pt]
NGC2264 C & Red  &   $<$  0.20   &$<$8.5$\times10^{-6}$\\[10pt]
$\rho$ Oph A & Blue &   0.80 (0.02) &\phantom{$<$}5.5$\times10^{-5}$\\
& Red  &   0.83 (0.02) &\phantom{$<$}3.6$\times10^{-5}$\\[10pt]
L1689N & Blue &   0.27 (0.05) &\phantom{$<$}9.0$\times10^{-6}$\\
& Red  &   $<$  0.15   &$<$8.1$\times10^{-6}$\\[10pt]
Ser SM1 & Blue &   $<$  0.12   &$<$4.2$\times10^{-6}$\\
& Red  &   0.39 (0.04) &\phantom{$<$}9.0$\times10^{-6}$\\[10pt]
L1157 & Blue &   0.10 (0.03) &\phantom{$<$}2.3$\times10^{-5}$\\
& Red  &   0.11 (0.04) &\phantom{$<$}3.4$\times10^{-5}$\\[10pt]
L1228 & Blue &   0.31 (0.02) &\phantom{$<$}2.1$\times10^{-5}$\\
& Red  &   0.28 (0.02) &\phantom{$<$}3.3$\times10^{-5}$\\[10pt]
IC1396N & Blue &   $<$  0.13   &$<$2.0$\times10^{-5}$\\
& Red  &   $<$  0.13   &$<$2.6$\times10^{-5}$\\[10pt]
S140 & Blue &   0.88 (0.03) &\phantom{$<$}1.0$\times10^{-5}$\\
& Red  &   0.37 (0.02) &\phantom{$<$}9.2$\times10^{-6}$\\[10pt]
Ceph A HW2 & Blue &   1.36 (0.04) &\phantom{$<$}2.2$\times10^{-5}$\\
& Red  &   1.36 (0.04) &\phantom{$<$}2.3$\times10^{-5}$\\[10pt]
\enddata
\tablenotetext{a} {The numbers in parantheses are 1-$\sigma$ statistical uncertainties on
the integrated intensity, line flux and observed isotopic ratio.  All limits quoted are 3-$\sigma$}
\tablenotetext{b}{CI abundances relative to H$_2$ assuming a H$_2$/CO ratio of 10$^4$}
\end{deluxetable}

\begin{figure}[hbt!]
\centering
\figurenum{1}
\plotone{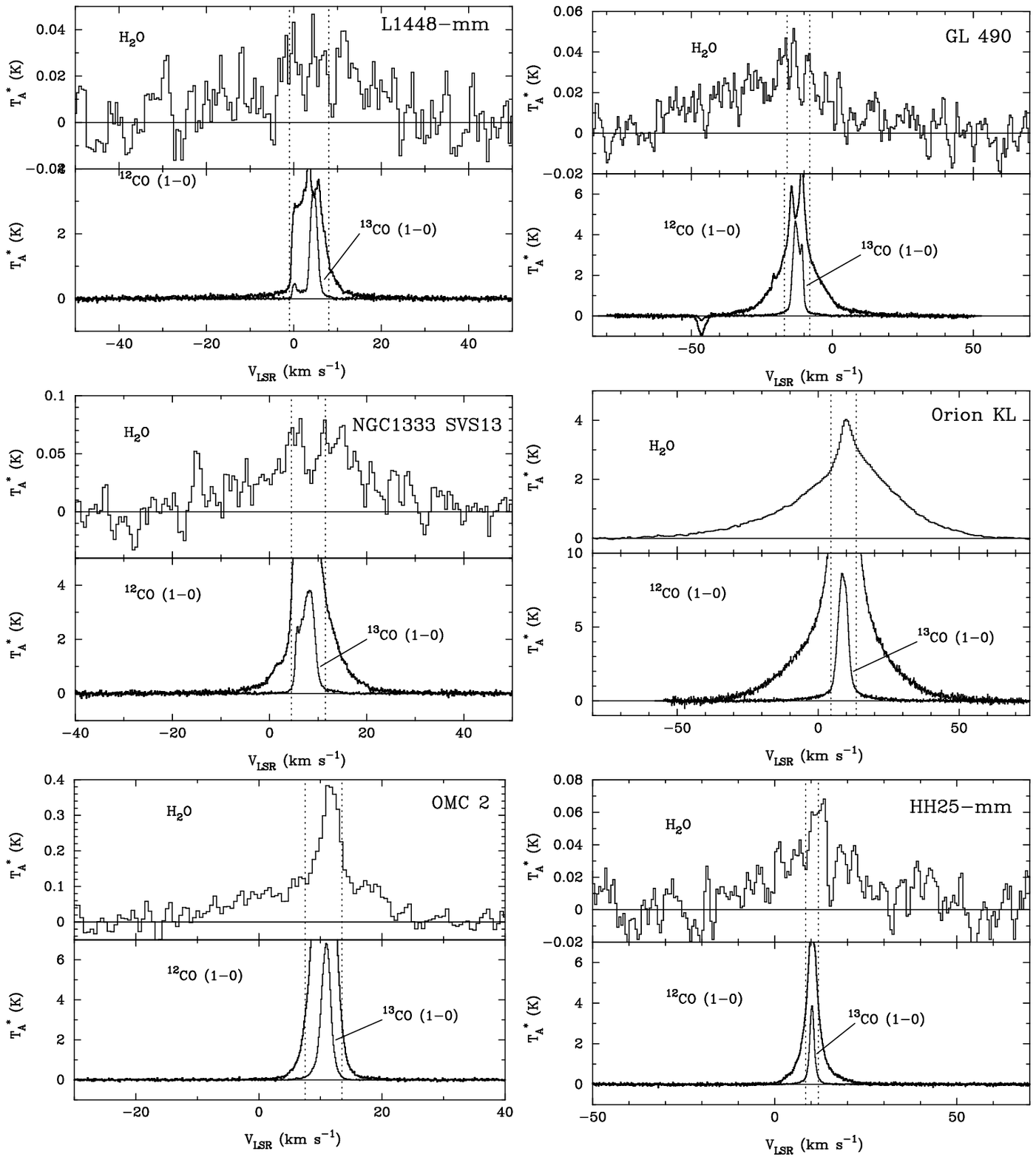}
\caption[]
{(a) Spectra towards L1448-mm, GL490, NGC1333-SVS13, OMC 1, OMC 2 and HH25-mm. The positions
observed are provided in Table 1.  For each source there are two panels, 
the upper panel shows the spectrum of the $1_{10}\rightarrow1_{01}$ transition of ortho-\h2o\
obtained with SWAS and the lower panel shows the spectra of the J=1$\rightarrow$0 transition of 
CO and the J=1$\rightarrow$0 transition of $^{13}$CO obtained at FCRAO spatially averaged in the
manner described in text.}
\label{}
\end{figure}

\begin{figure}[hbt!]
\centering
\figurenum{1}
\plotone{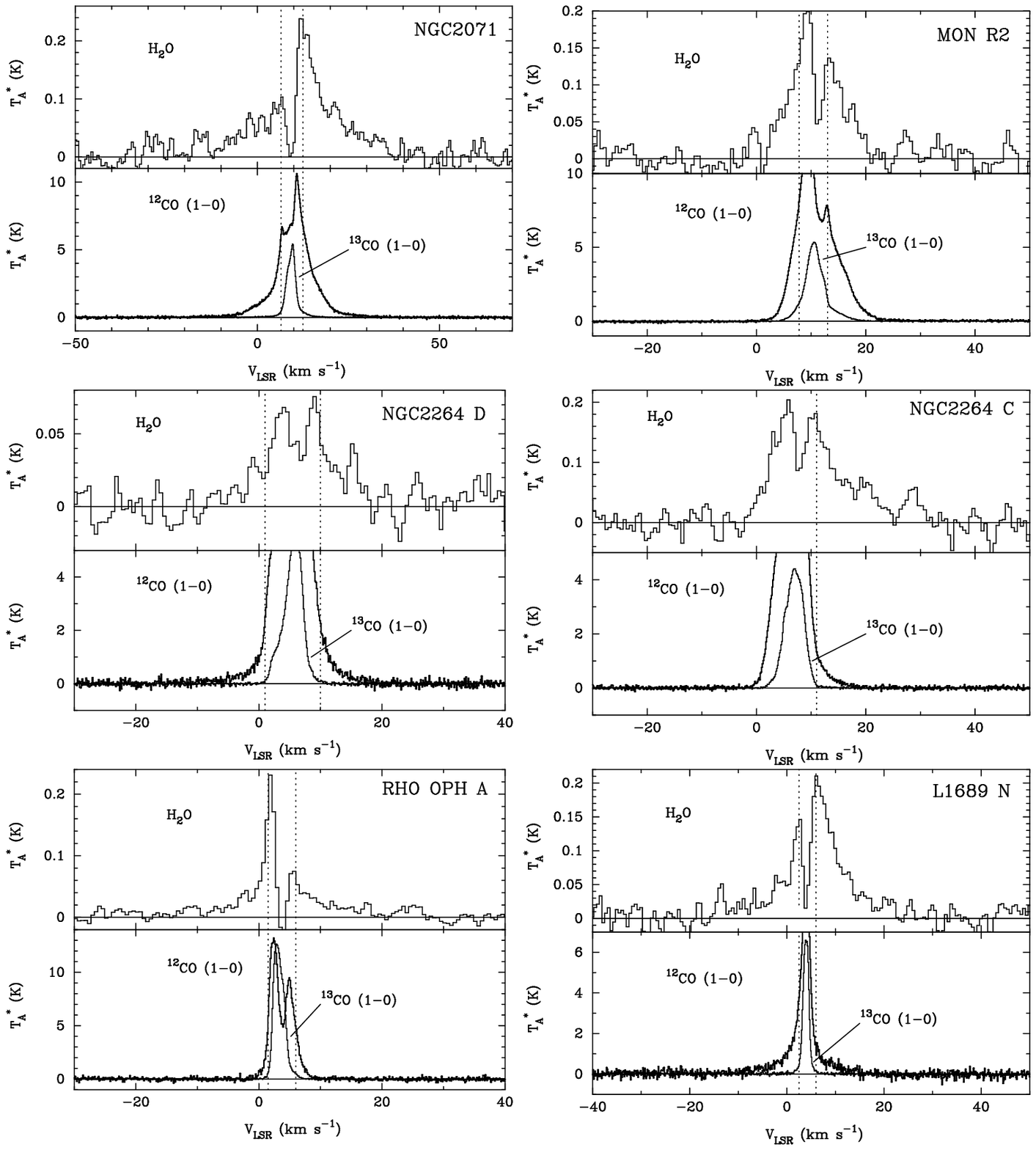}
\caption[]
{(b) The same as Fig. 1a for NGC2071, MonR2, NGC2264 D, 
NGC2264 C, $\rho$ Oph A, and L1689 N. }
\label{}
\end{figure}

\begin{figure}[hbt!]
\centering
\figurenum{1}
\plotone{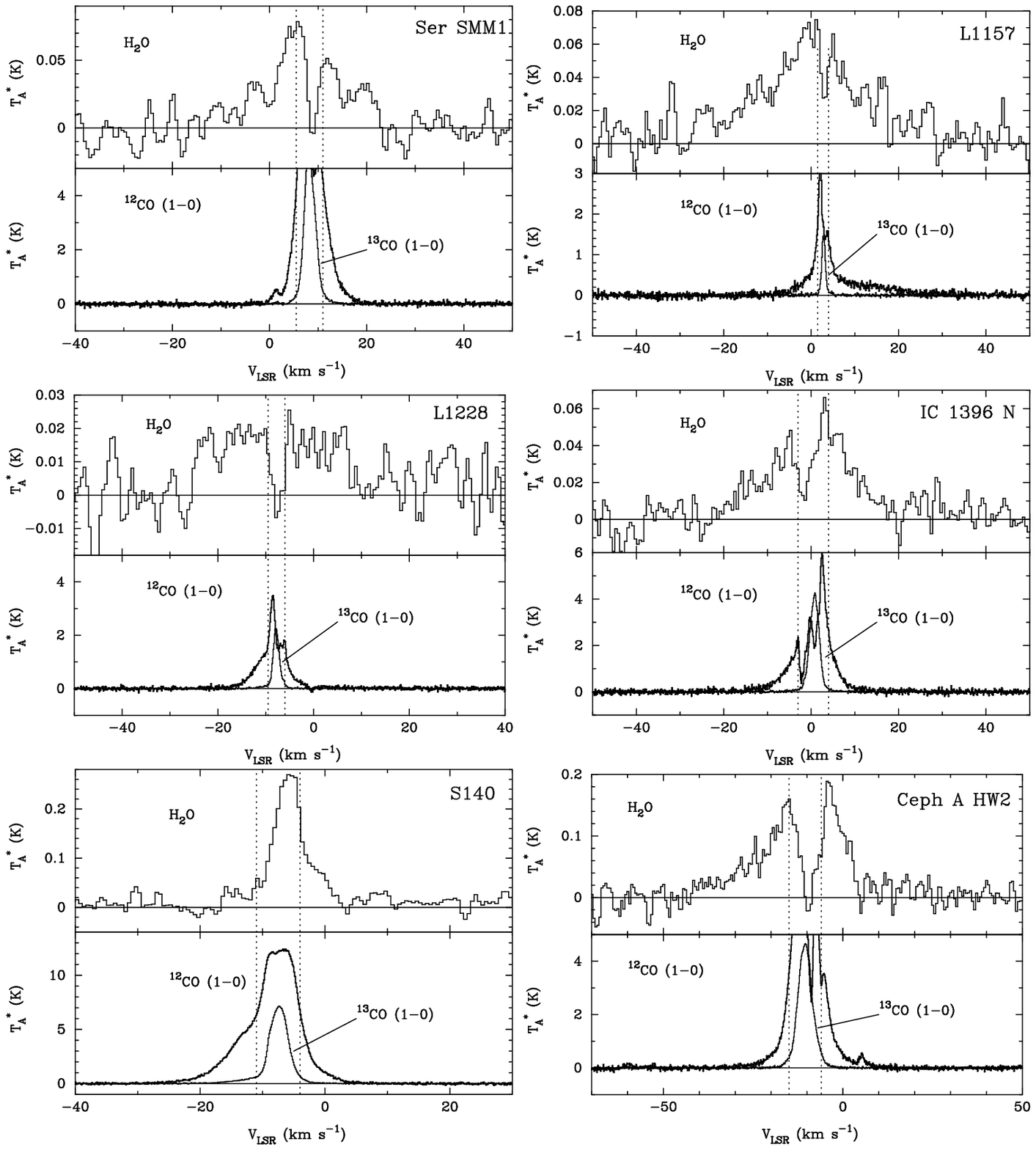}
\caption[]
{(c) The same as Fig. 1a for Ser SMM1, L1157, L1228, 
IC 1396N, S140 and Ceph A HW2.}
\label{}
\end{figure}

\begin{figure}[hbt!]
\centering
\figurenum{2}
\plotone{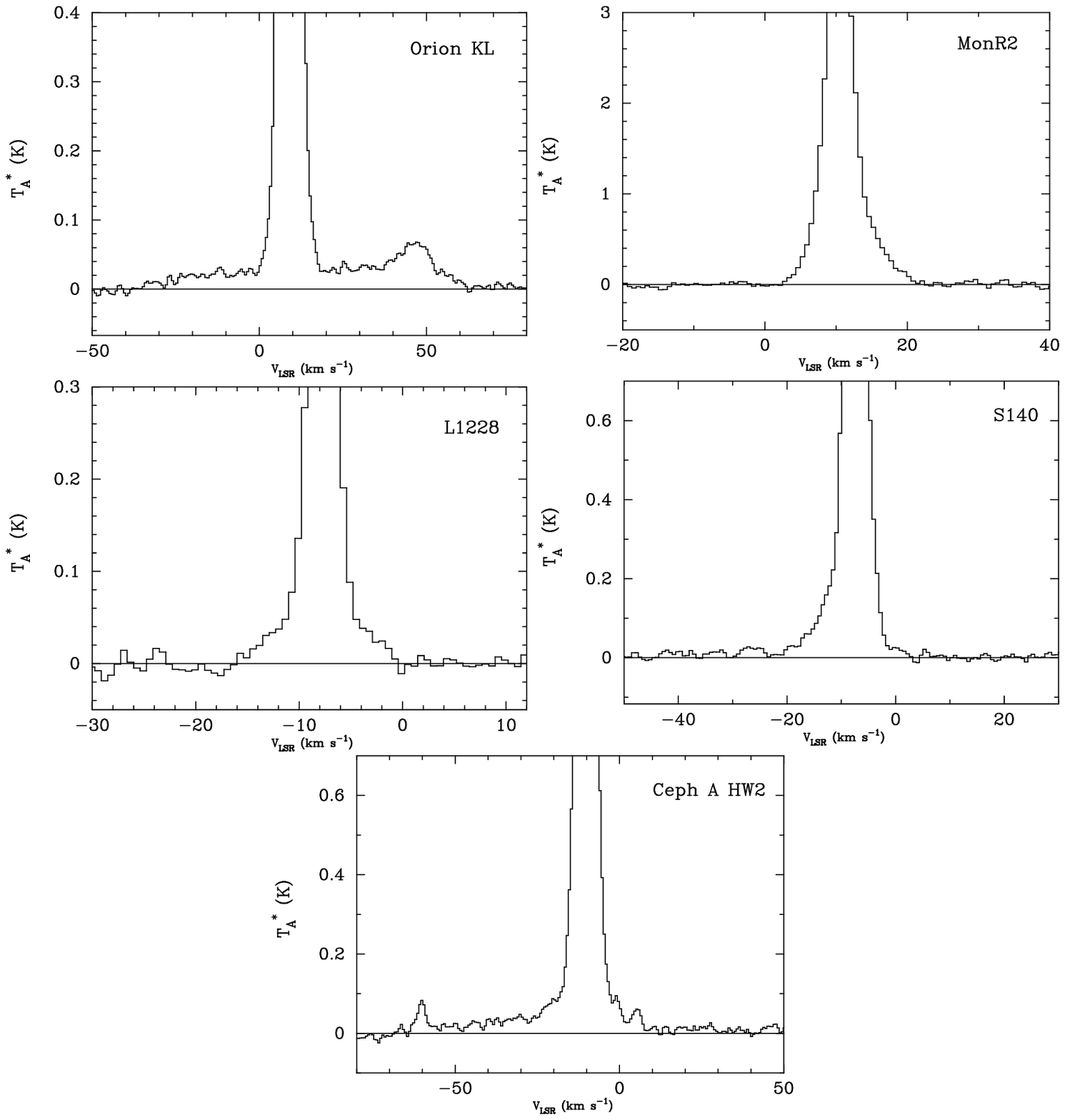}
\caption[]
{Spectra of the $^3P_1 \rightarrow\ ^3P_0$ transition of [CI] towards OMC 1, Mon R2,
L1228, S140 and Ceph A HW2 obtained with SWAS. The positions observed are provided in Table 1.}
\label{}
\end{figure}

\end{document}